\documentstyle[prl,aps,epsf]{revtex}
\begin{document}

\title{Square-well solution to the three-body problem}
\author{A.S.~Jensen and  E.~Garrido \\
Institute of Physics and Astronomy, \\
Aarhus University, DK-8000 Aarhus C, Denmark \\ and \\ 
D.V.~Fedorov \\ European Centre for Theoretical
Studies in Nuclear Physics \\ and Related
Areas,  I-38050 Trento, Italy \\ and \\
Institute of Physics and Astronomy, \\
Aarhus University, DK-8000 Aarhus C, Denmark }
\date{today}

\maketitle

\begin{abstract}
The angular part of the Faddeev equations is solved analytically for
s-states for two-body square-well potentials. The results are, still
analytically, generalized to arbitrary short-range potentials for both
small and large distances. We consider systems with three identical
bosons, three non-identical particles and two identical spin-1/2
fermions plus a third particle with arbitrary spin.  The angular
wave functions are in general linear combinations of trigonometric and
exponential functions. The Efimov conditions are obtained at large
distances. General properties and applications to arbitrary potentials
are discussed. Gaussian potentials are used for illustrations. The
results are useful for numerical calculations, where for example large
distances can be treated analytically and matched to the numerical
solutions at smaller distances. The saving is substantial.  \\
PACS numbers 3.65.Ge, 2.60.Nm, 11.80.Jy, 21.45.+v
\end{abstract}

\section{Introduction}
A new method to study the quantum mechanical three-body problem was
recently formulated \cite{fed93}. The first significant advantage is
the precise treatment of large distances where the correct asymptotic
behavior of the wave function is incorporated. Another advantage is the
parallel treatment of bound and continuum states, which in combination
with the long-range treatment opens the interesting possibility of
approaching the celebrated three-body Coulomb problem in the
continuum. The method solves the coordinate space Faddeev equations in
two steps. First the angular (and most difficult) parts of the
three-component wave functions are calculated by elaborate use of the
analytical knowledge of the large-distance behavior. Then the coupled
set of effective radial equations is solved numerically \cite{fed94b}.

The new method is designed to solve the Faddeev equations in
coordinate space with the advantage of an immediate intuitive
interpretation of the physics involved. (It may be considered as a
development from the approach used to study the properties of $H^-$
\cite{mac68}.) The advantages are seen in the analytical and
numerical treatment of extreme cases like the so-called Efimov states
\cite{fed93,efi70}, which has been suggested and looked for in both
molecules \cite{cor86} and nuclei \cite{fed94c}. These states occur
when at least two of the two-body subsystems simultaneously have an
s-state at zero energy arising from short-range interactions.  The
resulting infinitely many bound three-body states of $0^+$ nature are
extremely extended in space and extremely weakly bound. The long range
Coulomb interaction between only one pair of particles destroys the
effect.

The large-distance coupling takes place only between s-states in
different relative two-body subsystems \cite{fed94b}. Furthermore,
low-lying three-body states often contain large components of such
s-states. Different treatments are usually needed when long-range
interactions like the Coulomb potential are involved and we shall only
consider short-range interactions. We shall furthermore neglect the
coupling to higher angular momenta and confine ourselves to relative
s-states. The purpose of this paper is to classify the lowest
eigenvalues and describe how the solutions for small as well as large
distances can be obtained analytically for the angular part of the
Faddeev equations.  These solutions turn out to be exact for
square-well potentials where we also can find exact solutions at
intermediate distances. Thus we shall derive a semianalytic s-state
square-well solution to the Faddeev equations.

The solutions describe the Efimov anomaly and how the infinitely many
states continuously appear and disappear as function of the parameters
of the potential. The power of the method is especially due to the
analytical treatment of large distances which must be treated with
particular care for loosely bound quantum systems in low angular
momentum states. Also the connection between two- and three-body
large-distance behavior can be explored. Using the analytical results
for small and large distances in numerical calculations for arbitrary
potentials improve both precision and computational speed and enable
thereby investigations of otherwise inaccessible problems.

Other methods are available for investigations of the three-body
problem for baryons \cite{ric92}, for Borromean systems
\cite{zhu93,fed94a}, for molecular systems \cite{coo93,jac94} as well
as for Coulomb interacting particles \cite{ros91,lin95}. The results
from all these methods are needed in studies of the properties of a
variety of different systems of interest in physics, see for example
\cite{efi90,kie94,fri95,han95}. Some of the methods solve the
Schr\"{o}dinger equation directly, but the Faddeev equations are
needed to describe all the subtle correlations. This is especially
clearly seen in systems where two- and three-body asymptotic behavior
are mixed \cite{fed94b}. Borromean systems, characterized by genuine
three-body asymptotics only, are relatively easy to handle, whereas
weakly bound systems, where both two and three-particle correlations
are essential, require very careful treatment of the large distances.

Analytic results are rare in quantum mechanics where the
Schr\"{o}dinger equation should be solved. The Faddeev equations
further complicate analytic analyses. However, so far at least one
exception exists for identical spinless particles interacting via
two-body harmonic oscillator potentials \cite{bar92}. This potential
is infinite at large distance and scattering states therefore cannot
be studied. Also the behavior for short-range potentials in general
are excluded.

In this paper we outline in section 2 the general theoretical
framework and in section 3 we solve the angular eigenvalue problem for
s-states for a system of three identical bosons and spin-independent
interactions.  In section 4 we generalize to systems of three
different particles and in section 5 we consider a system with two
identical spin-1/2 particles plus a third particle. In section 6 we
give numerical illustrations and indicate qualitatively how to
generalize the results.  Tedious mathematical derivations are
collected in appendices. Finally we give a summary and the conclusions
in section 7.

\section{Theoretical framework}
The intrinsic Hamiltonian of the three-body system is given by
\begin{equation} \label{e1} 
	H =\sum_{i=1}^{3} \frac{p_{i}^2}{2m_{i}} - \frac{P^2}{2M} +
        \sum_{i>j=1}^{3} V_{ij} \; , 
\end{equation}
where $m_{i}$, ${\bf r}_i$ and ${\bf p}_i$ are mass, coordinate and
momentum of the $i$'th particle, $V_{ij}$ are the two-body potentials,
$P$ and $M$ are respectively the total momentum and the total mass of
the system. We shall use the (three sets of) hyperspherical
coordinates which consist of one radial coordinate $\rho$
(hyperradius) and five generalized angles $\Omega_i$, where i$=1,2,3$.
The precise definitions are given in appendix A. One of these sets of
hyperspherical coordinates is in principle sufficient for a complete
description.  The volume element is given by $\rho^5{\rm d}\Omega{\rm
d}\rho$ where ${\rm d}\Omega=\sin^2 \alpha
\cos^2 \alpha {\rm d}\alpha {\rm d}\Omega_x {\rm d}\Omega_y$.

\subsection{General procedure}
The total wave function $\Psi$ of the three-body system is written as a
sum of three components $\psi^{(i)}$ which in turn for each $\rho$ are
expanded in a complete set of generalized angular functions:
\begin{equation} \label{e3}
\Psi= \sum_{i=1}^{3}  \psi^{(i)}  = \frac {1}{\rho^{5/2}}
	\sum_{i,n} f_n(\rho)  \Phi_n^{(i)}(\rho ,\Omega_i)
	\; ,
\end{equation}
where the radial expansion coefficients $f_n(\rho)$ are component
independent and $\rho^{-5/2}$ is the phase-space factor. These
wave functions satisfy the three Faddeev equations \cite{fed95}
\begin{equation} \label{e7}
(T-E)\psi^{(i)} +V_{jk} (\psi^{(i)}+\psi^{(j)}+\psi^{(k)})=0 \; ,
\end{equation}
where $E$ is the total energy, $T$ is the kinetic energy operator and
$\{i,j,k\}$ is a cyclic permutation of $\{1,2,3\}$. The Faddeev
equations may have non-trivial spurious solutions, where each
component is non-vanishing while the sum corresponding to the
Schr\"{o}dinger wave function is identically equal to zero. The
components of such solutions are eigenfunctions of the kinetic energy
operator with eigenvalues equal to the total energy.

The procedure is now for each $\rho$ to solve the eigenvalue problem
for the five dimensional angular part of the Faddeev operator:
\begin{equation} \label{e9}
 {\hbar^2 \over 2m}\frac{1}{\rho^2}\hat\Lambda^2 \Phi_n^{(i)} +V_{jk}
(\Phi_n^{(i)}+\Phi_n^{(j)}+\Phi_n^{(k)}) = {\hbar^2 \over
2m}\frac{1}{\rho^2} \lambda_n (\rho) \Phi_n^{(i)} \;  ,
\end{equation}
where $\hat\Lambda^2$ is the $\rho$-independent part of the kinetic
energy operator defined by
\begin{equation}\label{e11} 
      T \equiv T_{\rho}+{\hbar^2 \over 2m}\frac{1}{\rho^2}\hat\Lambda^2, \;
      T_{\rho}=-{\hbar^2 \over 2m}\left(
\rho^{-5/2}\frac{{\partial}^2}{{\partial}\rho^2}
\rho^{5/2}-\frac{1}{\rho^2} \frac{15}{4}\right) \; .
\end{equation}
Explicitly the generalized angular momentum operator $\hat \Lambda^2$
is given by
\begin{equation}  \label{e13}
	\hat \Lambda^2=-{1 \over \sin\alpha\cos\alpha}
      {{\partial}^2\over {\partial} \alpha^2} \sin\alpha \cos\alpha 
	+{\hat l_x^2 \over {\sin^2 \alpha}}
 	+{\hat l_y^2 \over {\cos^2 \alpha}} -4 
\end{equation}
in terms of an arbitrary $\alpha$-coordinate and the angular momentum
operators $\hat l_{x}^2$ and $\hat l_{y}^2$ related to the Jacobi
coordinates. The spurious states are characterized by angular
eigenvalues equal to those of the angular kinetic energy operator
$\hat \Lambda^2$, i.e.\ $K(K+4), K=0,1,2...$

Insertion of $\psi^{(i)}$ defined in eq.~(\ref{e3}) into
eq.~(\ref{e7}) then leads to the coupled set of ``radial''
differential equations
\begin{equation} \label{e15}
   \left(-\frac{\rm d ^2}{\rm d \rho^2}
   -{2mE\over\hbar^2}+ \frac{1}{\rho^2}\left( \lambda_n(\rho) +
  \frac{15}{4}\right) \right)f_n
   = \sum_{n'} \left(
   2P_{nn'}{\rm d \over\rm d \rho}
   + Q_{nn'}
   \right)f_{n'}  \;   ,
\end{equation}
with the functions $P$ and $Q$ defined by
\begin{equation}\label{e17}
   P_{nn'}(\rho)\equiv \sum_{i,j=1}^{3}
   \int d\Omega \Phi_n^{(i)\ast}(\rho,\Omega)
   {\partial\over\partial\rho}\Phi_{n'}^{(j)}(\rho,\Omega),\;
\end{equation}
\begin{equation}\label{e19}
   Q_{nn'}(\rho)\equiv \sum_{i,j=1}^{3}
   \int d\Omega \Phi_n^{(i)\ast}(\rho,\Omega) 
   {\partial^2\over\partial\rho^2}\Phi_{n'}^{(j)}(\rho,\Omega) \;  .
\end{equation}
The diagonal part of the $P$-matrix vanishes, i.e.\ $P_{nn}=0$.

\subsection{Angular eigenvalue equation}
It is convenient to show explicitly the spin dependence of the
wave function $\Phi_n^{(i)}(\rho ,\Omega_i)$ in eq.~(\ref{e3}), see
e.g.~\cite{fed95}. Let us consider s-waves only and assume that ${\bf s}$
is the intermediate spin obtained by coupling of the spins ${\bf s}_j$
and ${\bf s}_k$ of particles $j$ and $k$. The spin ${\bf s}$ is
afterwards coupled to ${\bf s}_i$ to give the total spin S of the
system. The total angular wave function for the i'th channel then
factorizes into the spatial part $\phi_{n,s}^{(i)}(\rho ,\Omega_i)$
and the spin dependent part $\chi_{n,s}^{(i)}$, i.e.~\
\begin{eqnarray}\label{e21}
\Phi_{n}^{(i)}(\rho,\Omega_i) & = & 
     \frac{1}{\sin{(2 \alpha_i)}}  \sum_{s}
\phi_{n,s}^{(i)}(\rho,\Omega_i) \chi_{n,s}^{(i)}   \;  ,
\end{eqnarray}
where we explicitly extracted the phase-space factor $\sin{(2 \alpha_i)}$.
Both $\phi$ and $\chi$ may depend on the intermediate coupling. 

The Faddeev components in eq.~(\ref{e7}) must be expressed in one
Jacobi coordinate set. For the s-waves the wave functions
$\phi_{n,s}^{(k)}$, which only depend on $\alpha _k$ and $\rho$, are
first expressed in terms of the i'th set of hyperspherical coordinates
and subsequently integrated over the four angular variables describing
the directions of ${\bf x}_k$ and ${\bf y}_k$. This amounts, still for
s-waves, to the substitution formally expressed by the operator
$R_{ik}$ defined by
\begin{equation}\label{e23}
R_{ik}  \left[\frac{\phi_{n,s}^{(k)}}{\sin(2\alpha _k)}\right] \equiv
\frac{1}{\sin(2\varphi _{j} )} \frac{1}{\sin(2\alpha _i)}
\int_{|\varphi _{j}-\alpha _i|}^{\pi/2 - |\pi/2-\varphi _{j} -\alpha _i|}
 \phi_{n,s}^{(k)}(\rho,\alpha _k)d\alpha _k \; ,
\end{equation}
where the angle $\varphi_{j}$ is given by the masses as
\begin{equation}\label{e25}
 \tan \varphi_{j}= \sqrt{m_j(m_1+m_2+m_3)\over m_i m_k} \; .
\end{equation}
It is closely related to the transformation angle $\varphi_{ik}$
used in appendix A and defined by
\begin{equation}\label{e24}
 \varphi_{ik}= (-1)^p \varphi_{j} = \arctan \left((-1)^p \sqrt{m_j(m_1+m_2+m_3)
  \over m_i m_k}\right) \; ,
\end{equation}
where $(-1)^p$ is the parity of the permutation $\{i,k,j\}$. 

Substituting eq.~(\ref{e21}) into eq.~(\ref{e9}) we obtain, after
multiplication from the left with $\chi_{n,s}^{(i)}$, the angular
eigenvalue equation
\begin{eqnarray}\label{e26}
  \left(
 - \frac{\partial^2 \phi_{n,s}^{(i)}
 (\rho,\alpha _i)}{\partial \alpha_i^2}
 + ( \rho^2 \langle \chi_{n,s}^{(i)} | v_i(\rho \sin{\alpha_i}) |
 \chi_{n,s}^{(i)} \rangle - \tilde{\lambda}_n(\rho))
  \phi_{n,s}^{(i)}(\rho,\alpha _i)  + \rho^2 {\sin(2\alpha _i)} 
  \right. \nonumber \\ 
   \times \left. \sum_{s^\prime s^\prime {^\prime}}
  \langle  \chi_{n,s}^{(i)} | v_i(\rho \sin{\alpha_i}) |
 \chi_{n,s^\prime {^\prime}}^{(i)} \rangle  
 \left(C_{s^\prime {^\prime}s^\prime}^{ij} 
R_{ij}\left[\frac{\phi_{n,s^\prime}^{(j)}}{{\sin(2\alpha _j)}}\right] + 
  C_{s^\prime {^\prime}s^\prime}^{ik} 
R_{ik}\left[\frac{\phi_{n,s^\prime}^{(k)}}{{\sin(2\alpha _k)}} \right] \right) 
 \right) = 0  \; ,
\end{eqnarray}
where $v_i(x)= V_{jk}(x/a_{jk}) 2m / \hbar^2$ with $a_{jk}$ defined
in appendix A, $\tilde{\lambda}_n(\rho) = \lambda_n(\rho)+4$ and the
coefficients $C_{ss^\prime}^{ik}$ expressing the overlap of the spin
functions are given by
\begin{equation}\label{e29}
 C_{ss^\prime}^{ik} = \langle \chi_{n,s}^{(i)} |
 \chi_{n,s^\prime}^{(k)} \rangle \; .
\end{equation}
These matrix elements are diagonal for $i=k$, i.e.\
$C_{ss^\prime}^{ii} = \delta_{ss^\prime}$ and symmetric, i.e.\
$C_{ss^\prime}^{ik} = C_{s^\prime s}^{ki}$.

When the potential is diagonal in spin-space we obtain the much
simpler set of equations
\begin{eqnarray}\label{e27}
  \left(
 - \frac{\partial^2 \phi_{n,s}^{(i)}
 (\rho,\alpha _i)}{\partial \alpha_i^2}
 + ( \rho^2 \langle \chi_{n,s}^{(i)} | v_i(\rho \sin{\alpha_i}) |
 \chi_{n,s}^{(i)} \rangle   -\tilde{\lambda}_n(\rho))
  \phi_{n,s}^{(i)}(\rho,\alpha _i)  + \rho^2 {\sin(2\alpha _i)}
  \right. \nonumber \\ 
    \left. \times \langle \chi_{n,s}^{(i)} | v_i(\rho \sin{\alpha_i}) |
 \chi_{n,s}^{(i)} \rangle  \sum_{s^\prime}^{ }
 \left(C_{ss^\prime}^{ij} R_{ij} \left[\frac{\phi_{n,s^\prime}^{(j)}}
 {{\sin(2\alpha _j)}} \right] + 
  C_{ss^\prime}^{ik} R_{ik}\left[ \frac{\phi_{n,s^\prime}^{(k)}}
 {{\sin(2\alpha _k)}}\right] \right)  \right) = 0  \; .
\end{eqnarray}
Eqs.~(\ref{e26}) and (\ref{e27}) constitute  sets of equations
obtained for i=1,2,3 and all possible (i-dependent) values of s. As
usual the values of $\{i,j,k\}$ must here be a permutation of
$\{1,2,3\}$. 

\subsection{Spin independent interactions}
When the interactions are independent of spin, each Faddeev component
must factorize into a spin-part and a spatial part. Furthermore, the
spin can be factorized out of eq.~(\ref{e9}) and the structure of the
Faddeev equations then remains unchanged for the spatial parts
alone. The spin dependent wave function must then be the same for all
three Faddeev components and the spatial parts of the wave function in
eq.~(\ref{e21}) can at most differ by a normalization constant, i.e.~\
\begin{equation}\label{e33}
  \phi_{n,s}^{(i)}(\rho,\Omega) \equiv b_{s}^{(i)} 
  \phi_{n}^{(i)}(\rho,\Omega) \; ,
\end{equation}
\begin{equation}\label{e35}
 \sum_{s} b_{s}^{(i)} \chi_{n,s}^{(i)} = \sum_{s} b_{s}^{(k)} \chi_{n,s}^{(k)} 
 \; ,  \;  \;  i,k=1,2,3.    \; 
\end{equation}

Multiplication of eq.~(\ref{e35}) by $\chi_{n,s}^{(i)}$ from the left
then gives
\begin{equation}\label{e37}
  b_{s}^{(i)} = \sum_{s^\prime} C_{ss^\prime}^{ik} b_{s^\prime}^{(k)} 
 \; , \; \; i,k=1,2,3.  \;
\end{equation}
These equations are not independent, since
\begin{eqnarray}\label{e41}
 \sum_{s^\prime}  C_{ss^\prime}^{ik} b_{s^\prime}^{(k)} = 
 \sum_{s^\prime}  C_{ss^\prime}^{il} b_{s^\prime}^{(l)} \; ,  \;
\end{eqnarray}
for all values of $s$, $k$ and $l$. This is easily seen by use of the
closure relation of eq.~(\ref{e29}), i.e.
\begin{equation}\label{e39}
 C_{ss^\prime}^{ik} = \sum_{s^\prime {^\prime}} C_{ss^\prime 
 {^\prime}}^{il} C_{s^\prime {^\prime} {s^\prime}}^{lk} \; , \; \; l=1,2,3 \; .
\end{equation}
Thus eq.~(\ref{e37}) is valid for all k, if it only holds for one of
the values of k.  If $b_{s}^{(3)}$ is arbitrarily chosen and
$b_{s}^{(1)}$ and $b_{s}^{(2)}$ calculated (i=1,2 and k=3) from
eq.~(\ref{e37}), we can see by using eqs.~(\ref{e39}) and (\ref{e41}),
that eq.~(\ref{e37}) is valid for all other values of i and k. Thus
any choice of $b_{s}^{(i)}$ for one value of i provides the same
spin-independent solution of the Faddeev equations. 

In the symmetric case when the three particles furthermore have equal
masses and spatial interactions, the wave functions $\Phi_{n}^{(i)}$
corresponding to the different Faddeev components are independent of
$i$.  Then eq.~(\ref{e21}) implies that
\begin{equation}\label{e31}
 \sum_{s} \phi_{n,s}^{(i)}(\rho,\Omega)  \chi_{n,s}^{(i)}
  = \sum_{s} \phi_{n,s}^{(k)}(\rho,\Omega)  \chi_{n,s}^{(k)} \; ,
\end{equation}
which by use of eq.~(\ref{e33}) directly leads to eq.~(\ref{e35})

Inserting eq.~(\ref{e33}) into eq.~(\ref{e27}) we obtain by use of
eq.~(\ref{e37}) the equation
\begin{eqnarray}\label{e43}
 &  & \left(
 - \frac{\partial^2 \phi_{n}
 (\rho,\alpha _i)}{\partial \alpha_i^2}
 + ( \rho^2 v_i(\rho \sin{\alpha_i})-\tilde{\lambda}_n(\rho))
  \phi_{n}(\rho,\alpha _i)  \right.  \nonumber \\ 
   & & + \left. \rho^2 {\sin(2\alpha _i)} v_i(\rho \sin{\alpha_i}) 
 \left( R_{ij} \left[\frac{\phi_{n}}{{\sin(2\alpha _j)}}\right] + 
 R_{ik} \left[\frac{\phi_{n}}{{\sin(2\alpha _k)}}\right] 
 \right)  \right) = 0  \; ,
\end{eqnarray}
which determines $\phi_n \equiv \phi_n^{(i)}$ for i=1,2,3.

\section{Three identical bosons for spin independent interactions}
In the symmetric case with spin independent interactions the remaining
single Faddeev equation eq.~(\ref{e43}) can by use of
eqs.~(\ref{e23}) and (\ref{e25}) explicitly be written as 
\begin{eqnarray}\label{e45}
-\frac{\partial^2\phi(\rho,\alpha)}{\partial\alpha^2}  & + & 
(\rho^2 v(\rho \sin\alpha)-\tilde{\lambda}(\rho)) \phi(\rho,\alpha) \nonumber\\
 & + & \frac{4}{\sqrt{3}} \rho^2 v(\rho \sin\alpha)
\int_{|\pi/3 - \alpha|}^{\pi/2 - |\pi/6 - \alpha|}
\phi(\rho,\alpha')d\alpha' = 0  \; ,
\end{eqnarray}
where we omitted the label ``n''. We shall only consider short-range
potentials and often further restrict ourselves to square wells. It is
most convenient first to solve exactly for the schematic square-well
potential and afterwards generalize as much as possible. The two-body
potential is then a step function
\begin{equation}\label{e47}
V(r) = - V_0 \Theta(r<R_0) \; 
\end{equation}
and the reduced potential in eq.~(\ref{e45}) is therefore another
stepfunction
\begin{equation}\label{e49}
  v(x) = - v_0 \Theta(x<X) \;  , 
\end{equation}
where $v_0=2mV_0/\hbar^2$ and $X=R_0/\sqrt{2}$ when the normalization
mass is equal to the mass of one the particles. The angular Faddeev
equation in eq.~(\ref{e45}) is now solved analytically in different
intervals corresponding to increasing values of $\rho$. The decisive
quantity for the square-well potential is $\rho \sin\alpha$, which
determines whether $v$ is finite or vanishes. When $\rho \le X$ we
have $v = - v_0$ for all values of $\alpha~\epsilon~[0,\pi/2]$.

\subsection{Short-distance behavior:
 $0 \le \rho \le R_0 / \protect \sqrt{2} \protect $} The potential is
now constant for all $\alpha$ and eq.~(\ref{e45}) simplifies to
\begin{eqnarray}\label{e51}
-\frac{\partial^2\phi(\rho,\alpha)}{\partial\alpha^2} - 
(\rho^2 v_0 + \tilde{\lambda}(\rho)) \phi(\rho,\alpha)
 = \frac{4}{\sqrt{3}} \rho^2 v_0
\int_{|\pi/3 - \alpha|}^{\pi/2 - |\pi/6 - \alpha|}
\phi(\rho,\alpha')d\alpha'   \; .
\end{eqnarray}
The differential equation, without the right hand side, has vanishing
solutions at $\alpha = 0$ of the form $\sin(k\alpha)$, which by
insertion on the right hand side of eq.~(\ref{e51}) easily is shown to
remain of this form if $k$ is an even integer, i.e.~\ $k = 2n$, where
$n$ is an integer. A solution to eq.~(\ref{e51}),
\begin{equation}\label{e55}
  \phi(\rho,\alpha) \propto \sin(2n\alpha)  \; ,
\end{equation}
is then by insertion found to correspond to the values of
$\tilde{\lambda}_n$ given by
\begin{equation}\label{e53}
 \tilde{\lambda}_n = 4 n^2- \rho^2 v_0 \left( 1 - \frac{2}{n} (-)^n 
 \frac{2}{\sqrt{3}} \sin(n\pi/3)\right)    \; .
\end{equation}

This wave function is furthermore also the solution for all potentials
at $\rho = 0$ where the corresponding eigenvalue $\tilde{\lambda}_n = 4n^2$
then is obtained by demanding a vanishing wave function at $\alpha =
\pi /2$.  Since the potential only enters in combination with
$\rho^2$, a perturbative solution to second order in $\rho^2$ is
easily obtained for potentials, which are finite at the origin, by
using $v(x) \approx v(0) = - v_0$. Thus eqs.~(\ref{e53}) and
(\ref{e55}) are also solutions for any potential in first order
perturbation theory provided the depth of the square-well potential is
replaced by the value $v(0)$ of the potential at $\rho = 0$.

The solution for $n=2$ is independent of $\rho$ and corresponds to the
trivial solution (identically zero) to the Schr\"{o}dinger
equation. For $n>2$ the solution is only an approximation, since
the higher angular momenta (neglected here) contribute to these states.

\subsection{Intermediate distances: $ \pi/3  \le \alpha_0 \le 
  \pi/2  ,   R_0 / \protect \sqrt{2} 
   \le \rho \le R_0 \protect \sqrt{2/3}  $} 
The potential now vanishes for $\alpha_0 \le \alpha \le \pi/2$ (region
II) where
\begin{equation}\label{e57}
 \alpha_0 = \arcsin(R_0 / \rho \sqrt{2} )    \;
\end{equation}
and it remains constant and finite for $0 \le \alpha \le
\alpha_0$. The solution to eq.~(\ref{e45}) in region II (vanishing
potential) is therefore
\begin{equation}\label{e59}
 \phi_{II} = A_{II} \sin \left((\alpha-\pi/2) \sqrt{\tilde\lambda} \right)
 = \frac{1}{2i} A_{II} \left( e^{i(\alpha-\pi/2)\sqrt{\tilde\lambda}} 
 - e^{-i(\alpha-\pi/2)\sqrt{\tilde\lambda}}
 \right)  \;  ,
\end{equation}
where we already explicitly selected the solution vanishing at $\alpha
= \pi /2$.

To proceed we divide the $\alpha$-space into subregions:
\begin{eqnarray}\label{e60}
  & A: &   0 \le \alpha \le \alpha_0 - \frac{\pi}{3}  \nonumber \\
  & B: &   \alpha_0 - \frac{\pi}{3} \le \alpha \le \frac{2\pi}{3} - \alpha_0 \\
  & C: &  \frac{2\pi}{3} - \alpha_0 \le \alpha \le \alpha_0  \nonumber \\
  & II: & \alpha_0   \le \alpha \le \pi/2
  \nonumber    \;  .
\end{eqnarray}

These regions are marked in fig.~1, which shows the integration limits
of $\alpha ^\prime$ as function of $\alpha$ for the integral in
eq.~(\ref{e45}). The coupling between the different regions can then
be seen and expressed in the set of equations:
\begin{eqnarray}\label{e61}
 -\frac{\partial^2\phi_A(\rho,\alpha)}{\partial\alpha^2} - 
 (\rho^2 v_0 + \tilde{\lambda}(\rho)) \phi_A(\rho,\alpha) 
 =  \frac{4}{\sqrt{3}} \rho^2 v_0 
\int_{|\pi/3 - \alpha|}^{\pi/3 + \alpha} \phi_C(\rho,\alpha')d\alpha' \; ,
\end{eqnarray}

\begin{eqnarray}\label{e63}
& - \frac{\partial^2\phi_C(\rho,\alpha)}{\partial\alpha^2} - 
(\rho^2 v_0 + \tilde{\lambda}(\rho)) \phi_C(\rho,\alpha) 
 =  \frac{4}{\sqrt{3}} \rho^2 v_0 \left(
\int_{|\pi/3 - \alpha|}^{\alpha_0 - \pi/3} \phi_A(\rho,\alpha')d\alpha'
  \nonumber
 \right.  \\
& \left.
 +  \int_{2\pi/3 - \alpha_0}^{2\pi/3 - \alpha} \phi_C(\rho,\alpha')d\alpha'
+ \int_{\alpha_0 - \pi/3}^{2\pi/3 - \alpha_0} \phi_B(\rho,\alpha')d\alpha'
 \right)  \; ,
\end{eqnarray}

\begin{figure}[t]
\epsfxsize=12cm
\epsfysize=7cm
\epsfbox[-30 -400 520 -50]{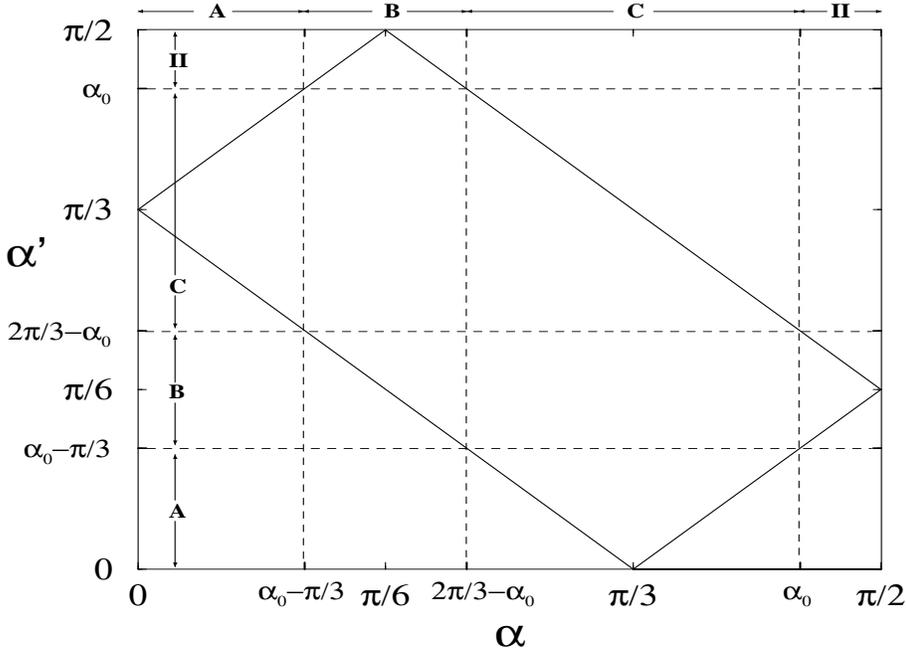}
\vspace{2.5cm}
\caption{\protect\small
The integration area in eq.~(\ref{e45}) is the
tilted rectangle (solid lines). The regions A,B,C and II (see
eq.~(\ref{e60})) are marked on the figure. The size of $\alpha_0$
corresponds to intermediate values of $\rho$: $R_0 / \sqrt{2} \le \rho
\le R_0 \sqrt{2/3}$. The thick solid line on the axis indicates the
range of $\alpha_0-$values.
}
\end{figure}

\begin{eqnarray}\label{e65}
& -\frac{\partial^2\phi_B(\rho,\alpha)}{\partial\alpha^2} - 
(\rho^2 v_0 + \tilde{\lambda}(\rho)) \phi_B(\rho,\alpha) 
 =  \frac{4}{\sqrt{3}} \rho^2 v_0 \left(
\int_{|\pi/3 - \alpha|}^{2\pi/3  - \alpha_0} \phi_B(\rho,\alpha')d\alpha'
 \nonumber 
 \right.  \\
& \left.
 + \int_{\alpha_0}^{\pi/2 - |\pi/6 - \alpha|} \phi_{II}(\rho,\alpha')d\alpha' +
\int_{2\pi/3 -  \alpha_0}^{\alpha_0} \phi_C(\rho,\alpha')d\alpha' \right)  \; ,
\end{eqnarray}
where the solutions are labeled according to subregion. Only $\phi_A$
and $\phi_C$ are directly coupled since $\phi_B$ enters in
eq.~(\ref{e63}) as an integral over a constant interval. Analogously
only $\phi_B$ and $\phi_{II}$ are directly coupled. 

Integrating the $\phi_{II}$-term in eq.~(\ref{e65}) we obtain
exponentially increasing and decreasing functions, exp$(\pm i\alpha
\sqrt{\tilde\lambda})$, which by further integration and differential
derivation still remain of the same functional form. Thus such
functions matching $\phi_{II}$ are necessary in the solution, but in
addition other exponentials are also possible. By insertion we then
find that the wave function
\begin{eqnarray}\label{e67}
 \phi_B = B_+^{II} e^{i\alpha \sqrt{\tilde\lambda}} + 
  B_-^{II} e^{-i\alpha \sqrt{\tilde\lambda}} 
  +  \sum_{k=1}^{3} 
  (B_+^{(k)} e^{\alpha \kappa_B^{(k)}} +  B_-^{(k)} 
 e^{-\alpha \kappa_B^{(k)}}) \;
\end{eqnarray}
is a solution to eq.~(\ref{e65}) when the B-coefficients and
$\kappa_B^{(k)}$ are related as shown in appendix B. We have unique
solutions for $B_+^{II}$ and $B_-^{II}$, except the patological case
of $\tilde\lambda =16/3$, and three different solutions for
$\kappa_B^{(k)}$ with three corresponding constraints between
$B_+^{(k)}$ and $B_-^{(k)}$. This is the explanation for the three
terms of the same form in eq.~(\ref{e67}). In addition there is a link
to the C-region providing one constraint between $\phi_C$ and the
$B_{\pm}^{(k)}$-values.

In the coupled A- and C-regions we must also look for exponential
functions as solutions. By insertion into eqs.~(\ref{e61}) and
(\ref{e63}) we find then that the wave functions
\begin{equation}\label{e69}
  \phi_A = \sum_{k=0}^{3} 
  A^{(k)} (e^{\alpha \kappa_{AC}^{(k)}} - e^{-\alpha \kappa_{AC}^{(k)}}) \;
\end{equation}
\begin{equation}\label{e71}
  \phi_C = \sum_{k=0}^{3} 
  (C_+^{(k)} e^{\alpha \kappa_{AC}^{(k)}} +  
 C_-^{(k)} e^{-\alpha \kappa_{AC}^{(k)}}) \;
\end{equation}
indeed are solutions provided the A- and C-coefficients and the
$\kappa_{AC}^{(k)}$-values are related as shown in appendix C. We have in
eq.~(\ref{e69}) already imposed the constraint that
$\phi_A(\alpha=0)=0$ which eliminated one of the coefficients present
in the other regions. We obtain four different solutions for
$\kappa_{AC}^{(k)}$ as indicated by the four k-values and the two
$C^{(k)}$-coefficients are uniquely determined by $A^{(k)}$. In addition there
is a link to the B-region providing one constraint between $\phi_B$
and the $A^{(k)}$-values.

The solutions are now explicitly written down in the four regions
named A,B,C and II. We have found wave functions containing the 8
parameters $A_{II}$, $B_-^{(k)}$, $A^{(0)}$ and $A^{(k)}$ for
k=1,2,3. Two linear constraints exist between them as seen from
eqs.~(B\ref{eb17}) and (B\ref{eb19}) in appendix B and
eqs.~(C\ref{ec17}) and (C\ref{ec19}) in appendix C. The matching
conditions at the three boundaries between the regions then provide
additional 6 linear constraints on the remaining 6 free
parameters. This leads as usual to the quantization condition for the
eigenvalue $\tilde\lambda$.

\begin{figure}[t]
\epsfxsize=12cm
\epsfysize=7cm
\epsfbox[-30 -400 520 -50]{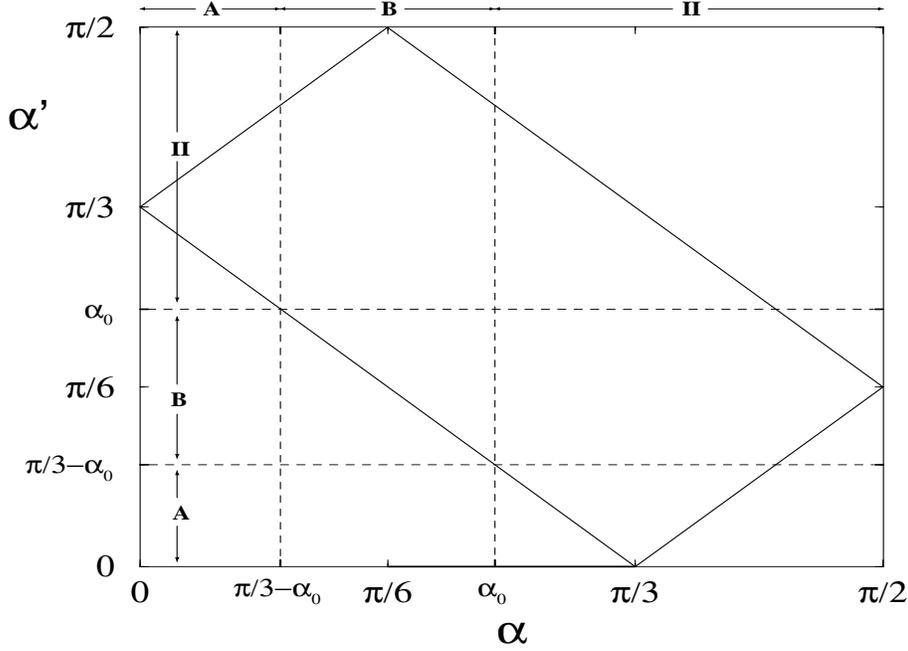}
\vspace{2.5cm}
\caption{\protect\small
The same as figure 1 for an $\alpha_0-$value
corresponding to intermediate values of $\rho$: $R_0 \sqrt{2/3} \le
\rho \le R_0 \sqrt{2}$. The regions marked on the figure are defined
in (see eq.~(\ref{e73})). The thick solid line on the axis indicates
the range of $\alpha_0-$values.
}
\end{figure}

\subsection{Intermediate distances: $ \pi/6  \le \alpha_0 \le 
  \pi/3 \; , \; 
R_0 \protect \sqrt{2/3} \protect \le \rho \le R_0 \protect \sqrt{2}
\protect $}

The potential again vanishes for $\alpha_0 \le \alpha \le \pi/2$,
where $\alpha_0$ is given in eq.~(\ref{e57}). Subregion C has now been
absorped into region II and subregions A and B are, as shown in fig.~2,
now defined by
\begin{eqnarray}\label{e73}
  & A: &   0 \le \alpha \le  \frac{\pi}{3} - \alpha_0  \nonumber \\
  & B: &   \frac{\pi}{3} - \alpha_0  \le \alpha \le \alpha_0 \\
  & II: &  \alpha_0   \le \alpha \le \pi/2
  \nonumber    \;  .
\end{eqnarray}
The coupled set of equations are now:
\begin{eqnarray}\label{e75}
 -\frac{\partial^2\phi_A(\rho,\alpha)}{\partial\alpha^2} - 
 (\rho^2 v_0 + \tilde{\lambda}(\rho)) \phi_A(\rho,\alpha) 
 =  \frac{4}{\sqrt{3}} \rho^2 v_0 
\int_{\pi/3 - \alpha}^{\pi/3 + \alpha} \phi_{II}(\rho,\alpha')d\alpha' \; ,
\end{eqnarray}
\begin{eqnarray}\label{e77}
& -\frac{\partial^2\phi_B(\rho,\alpha)}{\partial\alpha^2} - 
(\rho^2 v_0 + \tilde{\lambda}(\rho)) \phi_B(\rho,\alpha) 
 =  \frac{4}{\sqrt{3}} \rho^2 v_0 \left(
\int_{\pi/3 - \alpha}^{\alpha_0} \phi_B(\rho,\alpha')d\alpha'
 \nonumber 
 \right.  \\
& \left.
 + \int_{\alpha_0}^{\pi/2 - |\pi/6 - \alpha|} \phi_{II}(\rho,\alpha')d\alpha' 
 \right)  \; .
\end{eqnarray}
The couplings are much simpler and essentially only $\phi_B$ enters in
an integro-differential equation. Since $\phi_{II}$ still is given by
the expression in eq.~(\ref{e59}), a solution to eq.~(\ref{e75}) would
have to be proportional to $\sin(\alpha \sqrt{\tilde{\lambda}})$. The
solution, where the right hand side vanishes, is analogously
proportional to $\sin(\alpha \kappa)$, where
\begin{equation}\label{e91}
  \kappa \equiv  \sqrt{\rho^2 v_0 + \tilde{\lambda}(\rho)} \; .
\end{equation}
By insertion we then obtain the complete solution 
to eq.~(\ref{e75}) as
\begin{equation}\label{e79}
  \phi_A =  A_{f} (e^{i \alpha \sqrt{\tilde{\lambda}}} - 
        e^{-i \alpha \sqrt{\tilde{\lambda}}})
  + \frac{1}{2i} A_h (e^{i \alpha \kappa}    
    -  e^{-i \alpha \kappa})  = 
  2 i A_{f}  \sin \left(\alpha \sqrt{\tilde\lambda} \right)
  + A_h \sin (\alpha \kappa ) \; ,   
\end{equation}
where the coefficient $A_h$ is arbitrary and $A_{f}$ is given by
\begin{equation}\label{e81}
  A_{f} = \frac{2A_{II}}{\sqrt{3\tilde{\lambda}}}  
         (- e^{i \alpha \sqrt{\tilde{\lambda}}\;\pi/6} + 
        e^{-i \alpha \sqrt{\tilde{\lambda}}\;\pi/6}) =
    - \frac{4iA_{II}}{\sqrt{3\tilde{\lambda}}} 
  \sin \left(\alpha \sqrt{\tilde{\lambda}}\;\pi/6  \right) \; .
\end{equation}

The wave function in the B-region is a solution to eq.~(\ref{e77}),
which apart from a constant term arising from $\phi_C$ and the upper
limit of the integral of $\phi_B$, is identical to eq.~(\ref{e65}).
The solution is therefore given by eq.~(\ref{e67}), which by insertion
into eq.~(\ref{e77}) also in this case leads to the expressions for
the coefficients given in eqs.~(B\ref{eb1})$-$(B\ref{eb4}) in appendix
B. The only difference is the constants in eq.~(B\ref{eb5}), which now
is changed into
\begin{eqnarray}\label{e83}
&  \frac{A_{II}}{2\sqrt{\tilde\lambda}} 
 \left(e^{i \sqrt{\tilde\lambda}(\alpha_0 - \pi/2)} + 
 e^{-i \sqrt{\tilde\lambda}(\alpha_0 - \pi/2)}\right) + 
 \frac{1}{i \sqrt{\tilde\lambda}}
 \left(B_+^{II} e^{i \sqrt{\tilde\lambda}\alpha_0}
 - B_-^{II} e^{-i \sqrt{\tilde\lambda}\alpha_0} \right) \nonumber
 \\
&  + \sum_{k=1}^{3} \left[ \frac{1}{\kappa_B^k}
 \left(B_+^{(k)} e^{\kappa_B^{(k)} \alpha_0}
 - B_-^{(k)} e^{-\kappa_B^{(k)} \alpha_0} \right) \right]  = 0 \; .
\end{eqnarray}
With the expressions in eqs.~(B\ref{eb7}), (B\ref{eb9}) and
(B\ref{eb15}) for the coefficients we can rewrite the constraint from
eq.~(\ref{e83}) as
\begin{eqnarray}\label{e85}
 & \sum_{k=1}^{3}  \left[  \frac{B_-^{(k)}}{\kappa_B^{(k)}}
 \left(\pm i e^{\kappa_B^{(k)} (\pi/3 - \alpha_0)}
  - e^{- \kappa_B^{(k)} (2\pi/3 - \alpha_0)} \right) \right]
 \nonumber \\
&  +\frac{A_{II}}{2\sqrt{\tilde\lambda}} \left[
   (e^{i \sqrt{\tilde\lambda}(\pi/2 - \alpha_0)} 
  + e^{-i \sqrt{\tilde\lambda}(\pi/2 - \alpha_0)})
 \right.   \\
 & \left.
 +\frac{1}{1-\frac{16}{3\tilde\lambda}} \; \; \frac{4}{i \sqrt{3\tilde\lambda}}
  \left(e^{i \sqrt{\tilde\lambda}(\alpha_0 - \pi/6)}
  \left(1 - \frac{4}{i \sqrt{3\tilde\lambda}}\right) 
 -  e^{-i \sqrt{\tilde\lambda}(\alpha_0 - \pi/6)}
  \left(1 + \frac{4}{i \sqrt{3\tilde\lambda}}\right) \right) \right] = 0 
 \nonumber \; .
\end{eqnarray}

\begin{figure}[t]
\epsfxsize=12cm
\epsfysize=7cm
\epsfbox[-30 -400 520 -50]{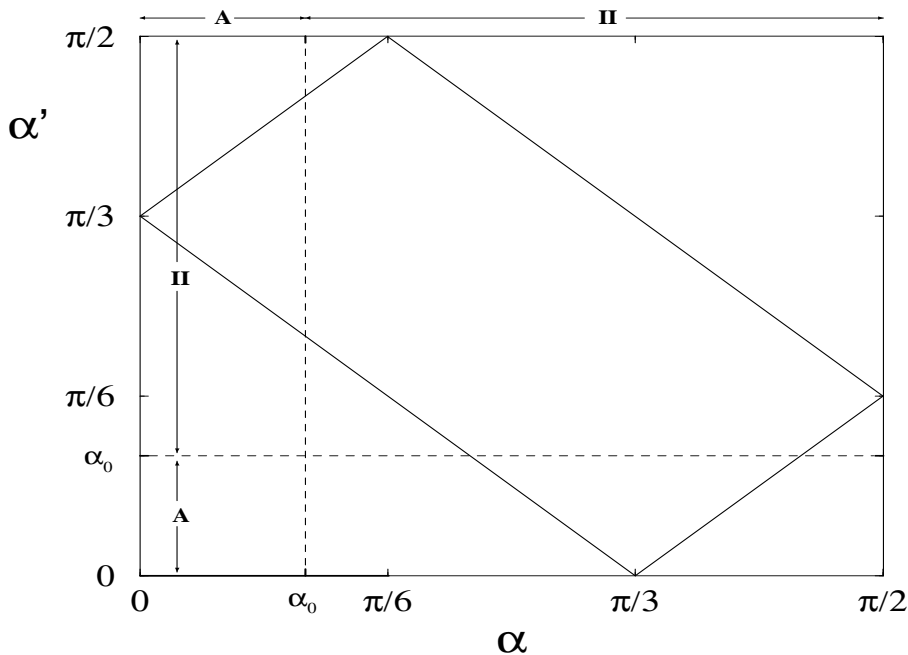}
\vspace{2.5cm}
\caption{\protect\small
The same as figure 1 for an $\alpha_0-$value
corresponding to large values of $\rho$: $R_0 \sqrt{2} \le \rho \le
\infty$. The regions marked on the figure are defined in (see
eq.~(\ref{e87})). The thick solid line on the axis indicates the range
of $\alpha_0-$values.
}
\end{figure}

We are now left with 5 parameters, i.e.~\ $A_{II}$, $A_h$ and
$B_-^{(k)}$ for k=1,2,3. One of these is determined by the constraint
in eq.~(\ref{e85}). The matching conditions at the two boundaries
between the regions provide additional 4 linear constraints on the
remaining 4 free parameters. Therefore we obtain again a quantization
condition for the eigenvalue $\tilde\lambda$.

\subsection{Large-distance behavior: $ 0 \le \alpha_0 \le 
  \pi/6 \; , \; 
R_0  \protect \sqrt{2} \protect \le \rho \le \infty$ }

The potential vanishes as usual for $\alpha_0 \le \alpha \le \pi/2$,
where $\alpha_0$ again is given in eq.~(\ref{e57}). Subregion B has
now also been absorped into region II and subregion A is, as
shown in fig.~3, now defined by
\begin{eqnarray}\label{e87}
  & A: &   0 \le \alpha \le \alpha_0   \nonumber \\
  & II: &  \alpha_0   \le \alpha \le \pi/2   \;  .
\end{eqnarray}
The equations now reduce to eq.~(\ref{e75}), which couples $\phi_A$
and $\phi_{II}$. The latter is again given by the expression in
eq.~(\ref{e59}) and consequently $\phi_A$ is given by eq.~(\ref{e79})
with one arbitrary coefficient and the other coefficient expressed in
eq.~(\ref{e81}).

The matching conditions at $\alpha_0$ provide 2 linear constraints on
the remaining 2 free parameter and we obtain again a quantization
condition for the eigenvalue $\tilde\lambda$. An explicit expression
is most conveniently derived by rewriting the solution in
eq.~(\ref{e79}) as
\begin{equation}\label{e89}
  \phi_A = \frac{8A_{II}} {\sqrt{3\tilde{\lambda}}}
 \sin (\sqrt{\tilde{\lambda}} \; \pi/6 ) \sin (\alpha \sqrt{\tilde{\lambda}})
  +  A_h \sin ( \alpha \kappa ) \; .
\end{equation}
Matching the wave functions in eqs.~(\ref{e59}) and (\ref{e89}) and
their first derivatives at $\alpha = \alpha_0$ gives immediately 
\begin{equation}\label{e93}
 \frac{8A_{II}}{\sqrt{3\tilde{\lambda}}}
  \sin( \sqrt{\tilde{\lambda}}\; \pi/6 ) \sin( \alpha_0 \sqrt{\tilde{\lambda}})
  +  A_h \sin( \alpha_0 \kappa ) 
   = A_{II} \sin \left((\alpha_0 - \pi/2) \sqrt{\tilde\lambda} \right) \;
\end{equation}
\begin{equation}\label{e95}
 \frac{8A_{II}}{\sqrt{3}}
  \sin( \sqrt{\tilde{\lambda}}\; \pi/6 ) \cos( \alpha_0 \sqrt{\tilde{\lambda}})
  +  A_h \kappa \cos( \alpha_0 \kappa ) 
   = A_{II} \sqrt{\tilde\lambda} 
 \cos \left((\alpha_0 - \pi/2) \sqrt{\tilde\lambda} \right) \; .
\end{equation}
The determinant $D$ corresponding to this linear set of equations for
$A_{II}$ and $A_h$ is
\begin{eqnarray}\label{e97}
& D =  \frac{\kappa}{\sqrt{\tilde{\lambda}}}  \cos( \alpha_0 \kappa)
  \left[ \frac{8}{\sqrt{3}} \sin( \sqrt{\tilde{\lambda}} \; \pi/6 ) 
  \sin( \alpha_0 \sqrt{\tilde{\lambda}}) - \sqrt{\tilde\lambda}
 \sin \left((\alpha_0 - \pi/2) \sqrt{\tilde\lambda} \right) \right]
   \nonumber \\
& - \sin(\alpha_0 \kappa)
 \left[ \frac{8}{\sqrt{3}} \sin( \sqrt{\tilde{\lambda}} \; \pi/6 ) 
  \cos( \alpha_0 \sqrt{\tilde{\lambda}}) - \sqrt{\tilde\lambda}
 \cos \left((\alpha_0 - \pi/2) \sqrt{\tilde\lambda} \right) \right] \; .
\end{eqnarray}
The non-trivial solutions are obtained for $D=0$, which therefore is
the eigenvalue equation for $\tilde\lambda$. 

We can simplify even further in the limit of very large distances, where
\begin{equation}\label{e99}
 \alpha_0 \approx \frac{R_0}{\rho\sqrt{2}} \; , \; \;
 \kappa \approx \rho \sqrt{v_0} \; , \; \;
 \kappa \alpha_0 \approx R_0 \sqrt{v_0/2} \; ,
\end{equation} 
where we assumed that $\tilde\lambda$ remains finite when $\rho$
increases towards infinity. The eigenvalue equation $D=0$ obtained
from eq.~(\ref{e97}) by expansion to the lowest orders in $1/\rho$
then becomes
\begin{equation}\label{e101}
 \frac{8}{\sqrt{3}} \sin( \sqrt{\tilde{\lambda}} \; \pi/6 )
 - \sqrt{\tilde\lambda} \cos (\sqrt{\tilde\lambda} \; \pi/2)
 =    \frac{\rho\sqrt{2}}{a_{scat}} \sin (\sqrt{\tilde\lambda} \; \pi/2)  \; ,
\end{equation}
where we introduced the two-body scattering length $a_{scat}$ of the
square-well potential
\begin{equation}\label{e103} 
  \frac{1}{a_{scat}} = \frac{\sqrt{v_0/2}}
  { - R_0 \sqrt{v_0/2} + \tan ({R_0 \sqrt{v_0/2}})}    \; .
\end{equation}
The factor of two appearing here is due to the mass (not the reduced
mass) in the definition of $v_0$. When $\rho$ approaches infinity
eq.~(\ref{e101}) can be valid only if $\tilde\lambda \rightarrow
4n^2$, where $n$ must be an integer. This is the same spectrum as
obtained in eq.~(\ref{e53}) for $\rho = 0$. Expanding to the next
order in $1/\rho$ gives
\begin{equation}\label{e104} 
  \tilde\lambda_n = 4n^2 \left(1 - \frac{12 a_{scat}}{\pi \rho \sqrt{2}}\right)
  \; .
\end{equation}

If the scattering length is infinitly large eq.~(\ref{e101}) reduces
to the equation derived by Efimov \cite{efi70}, which instead has the
solution $\tilde\lambda = -1.01251$ leading to infinitly many bound
states obtained from eq.~(\ref{e15}). Clearly if need be these
expressions can be extended to any order of correction in $1/\rho$.

When $\tilde\lambda / \rho^2$ diverges the asymptotic behavior is
different. We then define $\epsilon_x$ in the large-distance limit by
$\tilde\lambda \rightarrow 2m \epsilon_x \rho^2 / \hbar^2 \equiv
\bar{\epsilon_x} \rho^2$. It is easy to see that eq.~(\ref{e97}) at
most only can have solutions for negative $\epsilon_x$. Instead of
eq.~(\ref{e99}) we then at large $\rho$ obtain
\begin{eqnarray}\label{e90}
 \sqrt{\tilde\lambda} \approx -i\rho\sqrt{-\bar{\epsilon_x}} \; , \; \;
 \kappa \approx \rho \sqrt{v_0+\bar{\epsilon_x}} \; , \; \;
 \alpha_0 \sqrt{\tilde\lambda} \approx -i R_0 \sqrt{-\bar{\epsilon_x}/2}\; \; ,
 \nonumber \\ 
 \frac{\kappa}{\sqrt{\tilde\lambda}} \approx 
 i\sqrt{-(v_0+\bar{\epsilon_x})/\bar{\epsilon_x}}  \; \; , \; \;
 \kappa \alpha_0 \approx R_0 \sqrt{(v_0+\bar{\epsilon_x})/2} \; .
\end{eqnarray} 
Expanding eq.~(\ref{e97}) to lowest order in $1/\rho$ we find (instead
of eq.~(\ref{e101}))
\begin{equation}\label{e92}
 \tan \left( R_0 \sqrt{(v_0+\bar{\epsilon_x})/2} \right) \approx - \sqrt{
 \frac{v_0+\bar{\epsilon_x}}{-\bar{\epsilon_x}}} \; ,
\end{equation} 
which is the eigenvalue equation for the two-body subsystem with an
energy corresponding to $\bar{\epsilon_x}$. If two-body bound states
exist $\epsilon_x$ must be one of the energies. In other words for
each bound two-body state one $\tilde\lambda-$value diverges
parabolically according to eq.~(\ref{e90}).

For general short-range potentials, which vanish at distances larger
than a given finite radius $R_0$, we obtain again the
solutions described in this subsection. The Efimov conditions in
particular therefore remain the same for such potentials. 

\subsection{Large-distance behavior of $P_{nn^\prime}$ and $Q_{nn^\prime}$}
According to eqs.~(\ref{e17}) and (\ref{e19}) we must find the
derivative of the Schr\"{o}dinger wave function, $\Phi_n \equiv
\sum_{i=1}^{3} \phi_n^{(i)}$, with respect to $\rho$. In the present
symmetric case the three Faddeev components are equal. We choose to
work in the first Jacobi coordinate system and must then express the
orther two components in this set of coordinates. The transformation
is given in eq.~(\ref{e23}) and the angular wave function is
\begin{equation}\label{e201}
 \Phi \propto \phi_n(\alpha_1) + 2 R_{12}\left[\frac{\phi_n}
 {\sin(2\alpha _2)} \right] = \phi_n(\alpha_1) + \frac{4}{\sqrt{3}}
 \int_{|\pi/3 - \alpha _1|}^{\pi/2 - |\pi/6 -\alpha _1|}
 \phi_{n}(\rho,\alpha _2)d\alpha _2 \; ,
\end{equation}
where the factor 2 arises since the $R_{13}$-operation gives the same
result as that of $R_{12}$. The wave function $\phi_n(\alpha_1)$
explicitly is shown in eqs.~(\ref{e59}) and (\ref{e89}) with $A_h$
obtained from eq.~(\ref{e93}) as
\begin{equation}\label{e203}
 N(\sqrt{\tilde{\lambda}},\kappa,\alpha_0) \equiv \frac{A_h}{A_{II}} =
  \frac{1}{ \sin( \alpha_0 \kappa )} 
  \left[  \frac{8}{\sqrt{3\tilde{\lambda}}} 
  \sin( \sqrt{\tilde{\lambda}}\; \pi/6 ) \sin( \alpha_0 \sqrt{\tilde{\lambda}})
   - \sin \left((\alpha_0 - \pi/2) \sqrt{\tilde\lambda} \right) \right]  \; .
\end{equation}
It is then clear that $\Phi_n$ only depends on $\rho$ through
$\sqrt{\tilde{\lambda}_n},\kappa_n$ and $\alpha_0$. Thus
\begin{equation}\label{e205}
 {\partial\over \partial \rho}\Phi_{n} = 
 \frac{\partial \sqrt{\tilde{\lambda}_n}}
 {\partial \rho} \frac{\partial\Phi_{n}}{\partial\sqrt{\tilde{\lambda}_n}}
  +  \frac{\partial \kappa_n}
 {\partial \rho} \frac{\partial\Phi_{n}}{\partial\kappa_n}
  + \frac{\partial \alpha_0}
 {\partial \rho} \frac{\partial\Phi_{n}}{\partial\alpha_0} \; .
\end{equation}

For $P_{nn^\prime}$ we immediately obtain that
\begin{equation}\label{e207}
 P_{nn^\prime} \propto \frac{\partial \alpha_0} {\partial \rho}
 \int_0^{\pi/2} d\alpha_1 \Phi_n^{\ast}
   {\partial\over\partial\alpha_0}\Phi_{n'}  \; 
\end{equation}
as seen from the orthogonality of the different $\Phi_n$. Furthermore,
the derivative of the overall normalization factor $A_{II}$, which
also depends on $\alpha_0$, does not contribute to eq.~(\ref{e207})
again due to the orthogonality of the wave functions. The
large-distance behavior of ${\partial\alpha_0 /\partial \rho}$ is
seeen from the definition in eq.~(\ref{e57}) to be $\propto
1/\rho^2$.

We now divide the integration interval into the four parts appropriate for
the combination of the transformation between Jacobi coordinates
contained in eq.~(\ref{e201}) and the different wave functions given
in the two intervals of eq.~(\ref{e87}), i.e. 
\begin{equation}\label{e209}
 I_1=[0,\alpha_0] \; , \; I_2=[\alpha_0, \pi/3 - \alpha_0 ] \; , \;
   I_3=[\pi/3 - \alpha_0 , \pi/3 + \alpha_0] \; , \; 
 I_4=[\pi/3 + \alpha_0 , \pi/2] \; . \;
\end{equation}
As seen from eqs.~(\ref{e59}) and (\ref{e89}) and fig.3, the wave
function in eq.~(\ref{e207}) is apart from $A_{II}$ independent of
$\alpha_0$ in the intervals $I_2$ and $I_4$, where the contribution to 
$P_{nn^\prime}$ therefore is zero.

The third interval vanishes with $\alpha_0 \propto 1/\rho$ and the
$\alpha_0$-dependence of the wave function (again apart from $A_{II}$)
arises entirely from the $A_h$-term of the two transformed Faddeev
components. The derivative is proportional to $\kappa\alpha_0^2
\partial N / \partial \alpha_0$, which vanishes as $\alpha_0 \propto
1/\rho$. The contribution to $P_{nn^\prime}$ from the third interval
is then vanishing at least as fast as $1/\rho^4$.

The size of the first interval $I_1$ vanishes as $\alpha_0 \propto
1/\rho$. The $\alpha_0$-dependence of the wave function arises (apart
from $A_{II}$) entirely from the $A_h$-term in the first Faddeev
component. The derivative of $N$ is proportional to $\kappa\alpha_0$,
which remains finite together with the total wave function. Therefore
$P_{nn^\prime}$ in eq.~(\ref{e207}) vanishes as $1/\rho^3$ for all
$n,n^\prime$. The leading order term comes entirely from $I_1$, where
the transformed Faddeev components combined with the derivative of the
$A_h$-term of the first component contribute to the order $1/\rho^3$.

For the non-diagonal $Q_{nn^\prime}$ we also in analogy with
eq.~(\ref{e207}) obtain
\begin{eqnarray}\label{e211}
 Q_{nn^\prime} \propto \left(\frac{\partial \alpha_0} {\partial \rho}\right)^2
 \int_0^{\pi/2} d\alpha_1 \Phi_n^{\ast}
   {\partial^2\over\partial\alpha_0^2}\Phi_{n'} 
  + {\partial^2\alpha_0\over\partial\rho^2}
\int_0^{\pi/2} d\alpha_1 \Phi_n^{\ast}
   {\partial\over\partial\alpha_0}\Phi_{n'}  \; , 
\end{eqnarray}
where both terms therefore approach zero at least as $1/\rho^4$. The
first term due to $(\partial \alpha_0 / \partial \rho)^2  \propto 
1/\rho^4$ and the last term due to $\partial^2 \alpha_0 / \partial
\rho^2 \propto 1/\rho^3$ and the integral found in eq.~(\ref{e207})
vanishing as $1/\rho$.

The diagonal term $Q_{nn}$ can be written as a sum of six terms
corresonding to eq.~(\ref{e211}) and the analogous terms where
$\alpha_0$ is replaced by $\sqrt{\tilde{\lambda}_n}$ and
$\kappa_n$. The three terms with first order derivatives of the wave
function all vanish due to the normalization of $\Phi_n$. Since
$(\partial \alpha_0 / \partial \rho)^2 \propto 1/\rho^4$, we are left
with the leading terms
\begin{eqnarray}\label{e213}
 Q_{nn} \propto \left(\frac{\partial \kappa_n} {\partial \rho}\right)^2
 \int_0^{\pi/2} d\alpha_1 \Phi_n^{\ast}
   {\partial^2\over\partial\kappa_n^2}\Phi_{n} + 
 \left(\frac{\partial \sqrt{\tilde{\lambda}_n}} {\partial \rho}\right)^2
 \int_0^{\pi/2} d\alpha_1 \Phi_n^{\ast}
   {\partial^2\over\partial(\sqrt{\tilde{\lambda}_n})^2}\Phi_{n} \; ,
\end{eqnarray}
which for $n \ne n^\prime$ would vanish due to orthogonality. The
factor $(\partial \kappa_n /\partial \rho)^2$ approaches a constant
when $\rho \rightarrow \infty$ and so does $(\partial
\sqrt{\tilde{\lambda}_n} /\partial \rho)^2$ when
$\sqrt{\tilde{\lambda}_n}$ corresponds to a bound two-body state and
otherwise $(\partial \sqrt{\tilde{\lambda}_n} /\partial \rho)^2
\propto 1/\rho^4$. 

We again divide into the appropriate four intervals $I_1, I_2, I_3,
I_4$. Let us first study the case without two-body bound states where
the leading order term arises from the $\kappa$-derivation, since then
$(\partial \sqrt{\tilde{\lambda}_n} /\partial \rho)^2 \propto
1/\rho^4$. The second and fourth intervals contain no
$\kappa$-dependence and the contribution consequently vanish. The
third interval contributes at the most to the order $1/\rho^4$, i.e.
$\alpha_0 \propto 1/\rho$ from the interval size, $\alpha_0^2 \propto
1/\rho^2$ from the second derivative of the $A_h$-term of the two
transformed Faddeev components and finally $\alpha_0$ from the
interval size of the transformation integral. In the first interval we
have $\alpha_0 \propto 1/\rho$ from the size of the interval and
$\alpha_0^2 \propto 1/\rho^2$ from the second derivative of the
$A_h$-term in the first Faddeev component. The leading term of the
total wave function arises from the two tranformed Faddeev components
and it approaches a finite constant for $\rho \rightarrow \infty$.

The $\kappa_n$-dependence of the normalization constant $A_{II}$ did
not contribute to $P_{nn^\prime}$ due to orthogonality of the wave
functions. However, for $Q_{nn}$ we must consider this dependence. If
we define $\Phi_{n} \equiv A_{II} \tilde\Phi_{n}$, it is easy to show that
\begin{equation}\label{e215}
  \frac{\partial A_{II}}{\partial \kappa_n}   = 
  - |A_{II}|^3  \int_0^{\pi/2} d\alpha_1  \tilde\Phi_n^{\ast}
   {\partial\over\partial\kappa_n} \tilde\Phi_{n} \; .
\end{equation}
Going through the different intervals we find again vanishing
contributions from $I_2$ and $I_4$, a contribution proportional to
$1/\rho^3$ from $I_3$ and the leading order term $\propto 1/\rho^2$
from $I_1$ arising from the two transformed Faddeev components
combined with the derivative of the first Faddeev component.  We
therefore have $\partial A_{II} / \partial \kappa_n \propto 1/\rho^2$.
Furthermore, since
\begin{eqnarray}\label{e217}
 \int_0^{\pi/2} d\alpha_1 \Phi_n^{\ast}
 {\partial^2\over\partial\kappa_n^2}\Phi_{n} = 2 A_{II} 
 \frac{\partial A_{II}}{\partial\kappa_n} \int_0^{\pi/2} d\alpha_1  
 \tilde\Phi_n^{\ast} {\partial\over\partial\kappa_n} \tilde\Phi_{n} 
 \nonumber \\ 
 + A_{II} \frac{\partial^2 A_{II}}{\partial\kappa_n^2}
 \int_0^{\pi/2} d\alpha_1  \tilde\Phi_n^{\ast} \tilde\Phi_{n}
 + |A_{II}|^2 \int_0^{\pi/2} d\alpha_1  \tilde\Phi_n^{\ast} 
 {\partial^2\over\partial\kappa_n^2} \tilde\Phi_{n} \;
\end{eqnarray}
the terms with derivatives of $A_{II}$ approach zero at least as fast
as $1/\rho^3$ and so did the last term as discussed above.  Thus
$Q_{nn}$ approaches zero at least as fast as $1/\rho^3$.

In the case where $\tilde{\lambda}_n$ corresponds to a bound two-body
state, we get the same results for the same reasons as described
above.  However, then $(\partial \sqrt{\tilde{\lambda}_n} /\partial
\rho)^2$ approaches a constant for $\rho \rightarrow \infty$ and the
second term in eq.~(\ref{e213}) must also be considered. Although
possible this is a rather elaborate procedure due to the many terms
containing $\tilde{\lambda}_n$. We shall instead give arguments based
directly on the equation of motion.

At large distances the Faddeev components decouple as for example seen
from eq.~(\ref{e45}) where the potential in the last term vanishes
unless $\alpha$ approaches zero at least as fast as $1/\rho$. However,
when $\alpha \rightarrow 0$ the size of the integration interval
vanishes together with the integral itself. Thus the angular wave
function is in this limit determined by an ordinary Schr\"{o}dinger
equation, i.e.
\begin{equation}\label{e219}
 \left[-\frac{d^2}{dz^2} - \frac{\tilde\lambda}{\rho^2}
 + v(z) \right] \phi(z) = 0 \;
\end{equation}
where we introduced the new variable $z = \rho \sin \alpha \approx
\rho \alpha$. The boundary conditions $\phi(z=\rightarrow 0) = \phi(z
\rightarrow \infty) = 0$ make eq.~(\ref{e219}) equivalent to the
Schr\"{o}dinger equation for one of the (three identical) two-body
subsystems, i.e.
\begin{equation}\label{e221}
 \left[-\frac{d^2}{dz^2}  + v(z) \right] u(z) = 
  \frac{2m\epsilon_x}{\hbar^2} u(z) \;
\end{equation}
where $z=r\sqrt{m/\mu}$, $r$ is the relative coordinate, $\epsilon_x$
is the energy, $\mu$ the reduced mass of the two-body system.

From eqs.~(\ref{e219}) and (\ref{e221}) we obtain
\begin{equation}\label{e223}
 \tilde\lambda = \lambda + 4 = \frac{2m\epsilon_x}{\hbar^2} \rho^2  
  + \int dz u(z) \left[z \frac{du}{dz} + z^2 \frac{d^2u}{dz^2} \right] \;
\end{equation}
and $\phi(z)=\sqrt{\rho}u(z)$. Introducing $\phi(z)$ in the
$Q_{nn}$-equation we get
\begin{equation}\label{e225}
 Q_{nn} = -\frac{1}{4\rho^2}   
  + \frac{1}{\rho^2} \int dz u(z) \left[z \frac{du}{dz} 
  + z^2 \frac{d^2u}{dz^2} \right] \; .
\end{equation}
The integrals in eqs.~(\ref{e223}) and (\ref{e225}) cancel in the
combination $(\lambda + 15/4)/\rho^2 - Q_{nn}$ found in the coupled
set of radial equations, see eq.~(\ref{e15}). The result,
$2m\epsilon_x/\hbar^2$, restores the proper two-body radial
asymptotics which describes the motion of one of the particles against
the bound system of the other two particles.

We can conclude that the coupled set of radial equations in
eq.~(\ref{e15}) decouple in the limit of large $\rho$, since both
coupling terms $P_{nn^\prime}$ and $Q_{nn^\prime}$ approach zero at
least as fast as $1/\rho^3$, i.e. faster than that of the leading
centrifugal barrier term $1/\rho^2$. This behavior at large distances
is a general result valid for all short-range potentials.

\section{Three non-identical particles for spin independent interactions} 
In the asymmetric case with spin independent interactions the three
Faddeev equations in eq.~(\ref{e27}) can by use of eqs.~(\ref{e23}),
(\ref{e25}), (\ref{e33}) and (\ref{e37}) explicitly be written as
\begin{eqnarray}\label{e105}
& - \frac{\partial^2 \phi^{(i)}
 (\rho,\alpha _i)}{\partial \alpha_i^2}
 + ( \rho^2 v_i(\rho \sin{\alpha_i})-\tilde{\lambda}(\rho))
  \phi^{(i)}(\rho,\alpha _i)  + \rho^2 v_i(\rho \sin{\alpha_i})
 \left( \frac{1}{\sin(2\varphi _{j} )}   
 \right.  \\
& \left.    \times 
 \int_{|\varphi _{j}-\alpha _i|}^{\pi/2 - |\pi/2-\varphi _{j} -\alpha _i|}
 \phi^{(k)}(\rho,\alpha _k)d\alpha _k 
 + \frac{1}{\sin(2\varphi _{k} )}
  \int_{|\varphi _{k}-\alpha _i|}^{\pi/2 - |\pi/2-\varphi _{k} -\alpha _i|}
 \phi^{(j)}(\rho,\alpha _j)d\alpha _j \right)  = 0  \nonumber \; , \;
\end{eqnarray}
where $i=1,2,3$ and where we again omitted the label ``n''. We shall
follow the same procedure as in the symmetric case and mostly use
square-well potentials, i.e.~\
\begin{equation}\label{e107}
V_i(r) = - V_0^{(i)} \Theta(r<R_i) \; 
\end{equation}
and the reduced potentials in eq.~(\ref{e105}) are therefore the other
stepfunctions
\begin{equation}\label{e109}
  v_i(x) = - v_0^{(i)} \Theta(x<X_i) \;  , 
\end{equation}
where $v_0^{(i)}=2mV_0^{(i)}/\hbar^2$ and $X_i=R_i \mu_{jk}$. The
angular Faddeev equation in eq.~(\ref{e105}) are now solved
analytically in different intervals corresponding to increasing values
of $\rho$. The decisive quantity for each of the square-well
potentials is $\rho \sin\alpha_i$, which determines whether $v_i(x)$
is finite or vanishes.  When $\rho \le X_i$ we have $v_i(x) = - v_0^{(i)}$
for all values of $\alpha_i~\epsilon~[0,\pi/2]$.

Without loss of generality we can assume that $X_1 \le X_2 \le X_3$
which implies that $\alpha_0^{(1)} \le \alpha_0^{(2)} \le
\alpha_0^{(3)}$ where in analogy to eq.~(\ref{e57}) we define
\begin{equation}\label{e111}
 \alpha_0^{(i)} \equiv \arcsin(X_i / \rho)    \; .   
\end{equation}

We shall in this section confine ourselves to small or large
distances. The intermediate distances can be solved in analogy to the
symmetric case by division into $\rho$-intervals. The related
eigenfunctions are combinations of simple functions, but they are
rather tedious to write down in their full length and less interesting
as well.

\subsection{Short-distance behavior:  $0 \le \rho \le X_1 $}
The potentials are now all constants for all $\alpha$-values and
eq.~(\ref{e105}) is an inhomogeneous differential equation for
$\phi^{(i)}$, where the inhomogeneous part contains $\phi^{(j)}$ and
$\phi^{(k)}$. It can therefore be solved by adding one inhomogeneous
solution
\begin{equation}\label{e113}
  \phi^{(i)}(\rho,\alpha _i) = a_i \sin(2n \alpha _i)   \; 
\end{equation}
to the complete set of homogeneous solutions
\begin{equation}\label{e115}
  \phi^{(i)}(\rho,\alpha _i) = b_i \sin(\alpha _i \kappa_i)   \; \; , \;
  \kappa_i = \sqrt{v_0^{(i)}\rho^2 + \tilde{\lambda}(\rho)}   \; ,
\end{equation}
where we already imposed the boundary condition
$\phi^{(i)}(\rho,\alpha _i = 0) = 0$.

The inhomogeneous components in eq.~(\ref{e113}) must all be
proportional to $\sin(2n\alpha)$, since the integrated values in
eq.~(\ref{e105}) otherwise do not return the same function as required
by this equation.  The homogeneous components in eq.~(\ref{e115}) can
only be solutions if $v_0^{(1)}=v_0^{(2)}=v_0^{(3)}$ and $\varphi
_{1}=\varphi _{2}=\varphi _{3}=\pi/3$ which is the symmetric case
discussed in section~3. Any asymmetry therefore leads to
$b_1=b_2=b_3=0$.  A set of solutions is therefore obtained, if and
only if the three equations
\begin{eqnarray}\label{e117}
   \rho^2 (v_0^{(i)} - \epsilon) a_i =  \rho^2 v_0^{(i)} ( a_k d_j  + a_j d_k)
\end{eqnarray}
are fulfilled when $\{i,j,k\}$ are the three permutations of
$\{1,2,3\}$ and where we have defined
\begin{equation}\label{e120}
 \tilde{\lambda} \equiv 4 n^2- \rho^2 \epsilon \; \; \;  ,  \; \; \;
 d_i \equiv -\frac{\sin(2n\varphi_{i})}{n \sin(2\varphi_{i})} \; .
\end{equation}
In the limiting case of $\rho=0$ we immediately find the three times
degenerate solution $\tilde{\lambda}=4n^2$.

Non-trivial solutions for $\rho \neq 0$ only occur when the
corresponding determinant vanishes, i.e.~\
\begin{eqnarray}\label{e119}
 \left| \begin{array}{ccc}
  \epsilon -v_0^{(1)} &  v_0^{(1)} d_3 &  v_0^{(1)} d_2 \\
   v_0^{(2)} d_3 & \epsilon -v_0^{(2)} &  v_0^{(2)} d_1 \\
   v_0^{(3)} d_2 &  v_0^{(3)} d_1 & \epsilon -v_0^{(3)}
 \end{array}\right| 
   =  0  \;  ,
\end{eqnarray}
which determines the possible $\tilde{\lambda}(\rho)$. 

In the symmetric case ($v_0^{(1)}=v_0^{(2)}=v_0^{(3)} \equiv v_0$,
$d_1=d_2=d_3 \equiv d_0$) we obtain the two solutions
\begin{equation}\label{e121}
 \tilde{\lambda}(\rho) = 4 n^2- \rho^2 v_0 (1 + d_0)  \;
\end{equation}
\begin{equation}\label{e122}
  \tilde{\lambda}(\rho) = 4 n^2- \rho^2 v_0 (1 - 2 d_0) \; 
\end{equation}
where the first is two times degenerate and the last corresponds to
the symmetric solution as seen from eq.~(\ref{e117}).

In the general case we define
\begin{eqnarray}\label{e123}
 &  S_0  \equiv v_0^{(1)} v_0^{(2)} v_0^{(3)} (2d_1 d_2 d_3 + d_1^2 + d_2 ^2 + d_3^2 - 1)
  \nonumber \\
 &  S_1  \equiv v_0^{(1)} v_0^{(2)} (1 - d_3^2) + v_0^{(1)} v_0^{(3)} (1 - d_2 ^2)
   + v_0^{(2)} v_0^{(3)} (1 -d_1^2) \\
 & S_2  \equiv - v_0^{(1)} -v_0^{(2)} - v_0^{(3)}    \nonumber 
\end{eqnarray}
and rewrite eq.~(\ref{e119}) as 
\begin{equation}\label{e125}
 \epsilon ^3 + S_2 \epsilon ^2 + S_1 \epsilon + S_0  =0 \;  ,
\end{equation}
which is independent of $\rho$. Thus the short-distance behavior of
$\tilde{\lambda}(\rho)$ is quadratic in $\rho$ and given by
eq.~(\ref{e120}). 

For $n=1$ we have $d_i = -1$, $S_0 = S_1 = 0$ and consequently the
solutions $ \epsilon = -S_2$ and the doubly degenerate spurious solution
$\epsilon = 0$. This is also seen from eq.~(\ref{e121}) with $n=1$. The
corresponding wave function is determined by eq.~(\ref{e113}) with $a_i
\propto v_0^{(i)}$ as seen from eq.~(\ref{e117}).

For $n=2$ we have
\begin{equation}\label{e127}
 d_i = -\cos (2\varphi_{i}) = \frac{\tan^2\varphi_{i} - 1}
  {\tan^2\varphi_{i} + 1} \;
\end{equation}
and $S_0 =0$. Consequently we find the spurious solution $\epsilon = 0$
and the two solutions
\begin{equation}\label{e129}
 \epsilon = \frac{1}{2} \left(-S_2 \pm \sqrt{S_2^2 -4S_1} \right)  \; ,
\end{equation}
where $\epsilon$ is real, since $S_2^2 \ge 4S_1$ for all masses and
all values of the strength parameters $v_0^{(i)}$. The related
wave functions are found from eq.~(\ref{e117}).
  
For arbitrary $n$-values exceeding 2 we have three real solutions to
eq.~(\ref{e125}) as discussed in appendix D. For $n \rightarrow
\infty$, we have $d_i \rightarrow 0$.

Since the potentials also as in the symmetric case only enter in
combination with $\rho^2$, a perturbative solution to first order in
$\rho^2$ is obtained for arbitrary potentials by using $v_i(x) \approx
v_i(0) = - v_0^{(i)}$. Thus the solutions in this subsection are also
solutions for any potential in first order perturbation theory
provided the depth of the square-well potential is replaced by the
value $v_i(0)$ of the potential at $\rho = 0$.

\subsection{Large-distance behavior}
We define in this connection large distances to mean that
\begin{equation}\label{e131}
 \alpha_0^{(1)} \le \alpha_0^{(2)} \le \alpha_0^{(3)} \le 
  |\varphi_{1} - \alpha_0^{(1)}| \le |\varphi_{2} - \alpha_0^{(2)}| \le |\varphi_{3}
 - \alpha_0^{(3)}|  \;  
\end{equation}
for all sets ${1,2,3}$, which implies that $\alpha_0^{(i)} \le
\frac{1}{2}$Min$(\varphi_{i})$ for all sets ${1,2,3}$. The potentials
all vanish for $\alpha_i \ge \alpha_0^{(i)}$ and the solutions to
eq.~(\ref{e105}) are therefore the same as in eq.~(\ref{e59}), i.e.\
\begin{equation}\label{e133}
 \phi^{(i)}(\alpha_i) = a_i \sin \left((\alpha_i - \pi/2)
  \sqrt{\tilde\lambda} \right)  \; .
\end{equation}
In the other region where $\alpha_i \le \alpha_0^{(i)}$ we also obtain
decoupled solutions, i.e.\
\begin{equation}\label{e135}
  \phi^{(i)}(\alpha_i) = b_i \sin (\alpha_i \kappa_i)
 + c_i \sin \left(\alpha_i  \sqrt{\tilde\lambda} \right)  \;  ,
\end{equation}
where the first term is the solution to the homogeneous equation. The
constants $b_i$ are therefore completely free, since the functions to
be integrated in eq.~(\ref{e105}) only involve the functions in 
eq.~(\ref{e133}), i.e.\ values of $\alpha_i \ge \alpha_0^{(i)}$, i=1,2,3.

The connection between $a_i$ and $c_i$ are now found by insertion of 
eqs.~(\ref{e133}) and (\ref{e135}) into eq.~(\ref{e105}). The result is
\begin{eqnarray}\label{e137}
  \bar{c}_i =  \frac{2F^3}{\sqrt{\tilde\lambda}} (\bar{a}_j +  \bar{a}_k) \; ,
\end{eqnarray}
where
\begin{equation}\label{e136}
  \bar{a}_i = a_i / f_i  \; , \;   \bar{c}_i = - c_i f_i  \; , \; 
  f_i = \frac{\sin\left((\varphi_{i}-\pi/2)\sqrt{\tilde\lambda}\right)}
  {\sin(2\varphi_{i})}   \; , \;
 F =  (f_1 f_2 f_3)^{1/3}  \; .  
\end{equation}
This determines $c_i$ from given values of $a_i$.

The matching conditions at $\alpha_0^{(1)}, \alpha_0^{(2)}$ and $\alpha_0^{(3)}$
provide 6 linear constraints between the parameters $a_i$ and $b_i$,
where $c_i$ is found from eqs.~(\ref{e137})-(\ref{e136}). They only
have non-trivial solutions when the corresponding determinant
vanishes. In appendix E is shown that this condition, which determines
the eigenvalues $\tilde\lambda$, can be formulated as
\begin{equation}\label{e149}
 D = B_1 B_2 B_3 + 2A_1 A_2 A_3 - B_1 A_2 A_3  - A_1 A_2 B_3 -  A_1 B_2 A_3 
    = 0  \; ,
\end{equation}
where $A_i$ and $B_i$ are defined in appendix E.

In the symmetric case, ($B_1=B_2=B_3=B_0$, $A_1=A_2=A_3=A_0,
f_1=f_2=f_3$), we have $D = (A_0 - B_0)^2 (2A_0 + B_0)$, where the
solutions to $D=0$ correspond to those of eqs.~(\ref{e121}) and
(\ref{e122}).

Eq.~(\ref{e149}) simplifies in the limit of very large distances,
where
\begin{equation}\label{e151}
 \alpha_0^{(i)} \approx \frac{X_i}{\rho} \; , \; \;
 \kappa_i \approx \rho \sqrt{v_0^{(i)}} \; , \; \;
 \kappa_i \alpha_0^{(i)} \approx X_i \sqrt{v_0^{(i)}} \; ,
\end{equation} 
where we assumed that $\tilde\lambda$ remains finite. Expansion to lowest
order in $1/\rho$ then gives, see eqs.~(E\ref{ee5}) and (E\ref{ee11}),
\begin{equation}\label{e153}
 A_i \approx -2 F \mu_{jk} \sqrt{v_0^{(i)}} a_{scat}^{(i)} \cos(X_i\sqrt{v_0^{(i)}}) \; ,
\end{equation} 
\begin{equation}\label{e155}
 B_i \approx - \frac{f_i^2}{F^2}\cos(X_i\sqrt{v_0^{(i)}}) \sqrt{v_0^{(i)}}
 \left(\rho \sin(\frac{\pi}{2}\sqrt{\tilde\lambda}) + 
  a_{scat}^{(i)} \mu_{jk} \sqrt{\tilde\lambda}
  \cos(\frac{\pi}{2}\sqrt{\tilde\lambda}) \right) \; ,
\end{equation}
where the scattering length for the two-body system in analogy with
eq.~(\ref{e103}) is given by
\begin{equation}\label{e157}
  \frac{1}{a_{scat}^{(i)}} = \frac{\mu_{jk}\sqrt{v_0^{(i)}}}
  { - X_i \sqrt{v_0^{(i)}} + \tan \left({X_i \sqrt{v_0^{(i)}}}\right)}    \; .
\end{equation} 

The eigenvalues for $\rho \rightarrow \infty$ therefore approach
solutions to $\sin(\frac{\pi}{2}\sqrt{\tilde\lambda}) = 0$, i.e.\ the
hyperspherical spectrum of $\tilde\lambda = 4n^2$. The equation to the
next order in $1/\rho$ is instead
\begin{equation}\label{e159}
 \tan(\frac{\pi}{2}\sqrt{\tilde\lambda}) = - \frac{\sqrt{\tilde\lambda}}
  {\rho}  \sum_{i=1}^{3} a_{scat}^{(i)} \mu_{jk}  \; ,
\end{equation} 
which immediately  generalizes the symmetric case in eq.~(\ref{e104}) into 
\begin{equation}\label{e160}
  \tilde\lambda \approx 4n^2 ( 1 - \frac{4}{\pi \rho} 
  \sum_{i=1}^{3} a_{scat}^{(i)} \mu_{jk} )  \; .
\end{equation} 

The different patological cases of extremely large scattering lengths
are very different. Without loss of generality we can here
assume that $|a_{scat}^{(1)}| \leq |a_{scat}^{(2)}| \leq |a_{scat}^{(3)}|$ and
consider various cases. Using eqs.~(\ref{e149}), (\ref{e153}) and
(\ref{e155}) we obtain for large values of $\rho$ that the angular
eigenvalues when $|a_{scat}^{(2)}| \ll \rho \ll |a_{scat}^{(3)}|$ still are
given by $\sin(\frac{\pi}{2}\sqrt{\tilde\lambda}) = 0$ and when
$|a_{scat}^{(1)}| \ll \rho \ll |a_{scat}^{(2)}|$ instead are given by
\begin{equation}\label{e161}
  \pm \sqrt{\tilde\lambda} \cos(\frac{\pi}{2}\sqrt{\tilde\lambda})
  \sin(2\varphi_{1})  =  2 \sin\left((\varphi_{1}-\pi/2)\sqrt{\tilde\lambda}\right)
  \;  
\end{equation} 
and finally when $\rho \ll |a_{scat}^{(1)}|$ we find
\begin{equation}\label{e163}
 {\left(\frac{\sqrt{\tilde\lambda} \cos(\frac{\pi}{2}\sqrt{\tilde\lambda})}
  {2F}\right)}^3  - 
 \left(\frac{\sqrt{\tilde\lambda} \cos(\frac{\pi}{2}\sqrt{\tilde\lambda})}
  {2F}\right) \frac{(f_1^2 + f_2^2 + f_3^2)}{F^2}  + 2 = 0  \; .
\end{equation} 
The last equation has the symmetric solution, $\sqrt{\tilde\lambda}
\cos(\frac{\pi}{2}\sqrt{\tilde\lambda}) = - 4F$, when all the
quantities are independent of i. These results reflect the fact that
the Efimov anomaly only is present in the last two cases where at
least two of the scattering lengths are infinitly large with the
resulting infinitly many bound three-body states obtained from
eq.~(\ref{e15}). Clearly we are able to extend these expressions to
any order of correction in $1/\rho$.

When $\tilde\lambda / \rho^2$ diverges we can as in the symmetric case
again define $\epsilon_x$ in the large-distance limit by
$\tilde\lambda \rightarrow 2m \epsilon_x \rho^2 / \hbar^2 \equiv
\bar{\epsilon_x}\rho^2$.  Eq.~(\ref{e149}) can at most have solutions
for negative $\bar{\epsilon_x}$. At large $\rho$ we have approximately
\begin{eqnarray}\label{e162}
 \sqrt{\tilde\lambda} \approx -i\rho\sqrt{-\bar{\epsilon_x}} \; , \; \;
 \kappa_i \approx \rho \sqrt{v_0^{(i)}+\bar{\epsilon_x}} \; , \; \;
 \alpha_0^{(i)} \sqrt{\tilde\lambda} \approx -i X_i \sqrt{-\bar{\epsilon_x}/2}\; \; ,
 \nonumber \\ 
 \frac{\kappa_i}{\sqrt{\tilde\lambda}} \approx 
 i\sqrt{-(v_0^{(i)}+\bar{\epsilon_x})/\bar{\epsilon_x}}  \; \; , \; \;
 \kappa_i \alpha_0^{(i)} \approx X_i \sqrt{v_0^{(i)}+\bar{\epsilon_x}} \; .
\end{eqnarray} 
The dominating term in the determinat in eq.~(\ref{e149}) is
$B_1B_2B_3$ which diverges as $\rho^3$. To leading order in $\rho$ we
therefore obtain the eigenvalue solutions from $B_i=0$. This is
equivalent (see eqs.~(E\ref{ee5}) and (E\ref{ee11})) to $d_{ii}=0$
which immediately to leading order is seen to be identical to
eq.~(\ref{e92}) for the i'th two-body subsystem. Thus we obtain the
general result that there is a one to one correspondence between the
total number of two-body bound states in all the subsystems and the
parabolically diverging $\tilde\lambda-$values at large $\rho$.

The behavior of $P_{nn^\prime}$ and $Q_{nn^\prime}$ at large distance
is qualitatively the same as for three identical particles. This
result, as well as the various solutions described in this subsection,
is valid in general for all short-range potentials. In particular,
the Efimov conditions remain the same for such potentials.

\section{Two identical spin-1/2 particles} 
We consider a system of two identical spin--1/2 particles (labeled
``f'' and refered to as fermions) and a third particle (labeled by
``c'' and called the core) of spin $s_c$. The total angular momentum
$J$ can then take the values $J=s_c, s_c\pm 1$, since we only consider
vanishing orbital angular momentum.  The spin wave functions
$\chi_s^{(i)}$ from eq.~(\ref{e21}) are $\chi_0^{(1)}$, $\chi_{s_c \pm
1/2}^{(2)}$, and $\chi_{s_c \pm 1/2}^{(3)}$, which are defined as in
subsection 2.2. We have labeled the core as particle 1 and the two
fermions as particles  2 and 3.  The spin state $\chi_1^{(1)}$ is not
possible due to the required antisymmetrization of the wave function
under exchange of the fermions. Since we only include $s$-waves
the two fermions can not be coupled to spin 1. The total angular
momentum is therefore confined to be $J=s_c$.

The form of the Faddeev equations depends on the spin dependence of the
interactions. The fermion-fermion interaction, $V_{ff}(r)$, is only
effective in relative s-states and no spin dependence is needed. For
the fermion--core interaction we take the spin--dependence to be of the
form
\begin{equation}
V_2(r) = V_3(r) =  (1+\gamma_s \mbox{\bf s}_c \cdot \mbox{\bf s}_f)
     V_{fc}(r) \; .
\end{equation}
This potential is diagonal in spin space, and the diagonal matrix
elements for $i=2,3$ are:
\begin{eqnarray}
\langle \chi_{s_c+1/2}^{(i)}|V_{i}(r)|\chi_{s_c+1/2}^{(i)}\rangle 
 &\equiv& V_{fc}^+ (r) = 
   (1+ \gamma_s \frac{s_c}{2}) V_{fc}(r)  \\
 \langle \chi_{s_c-1/2}^{(i)}|V_{i}(r)|\chi_{s_c-1/2}^{(i)}\rangle 
 &\equiv& V_{fc}^- (r) = 
  (1- \frac{\gamma_s (s_c+1)}{2})  V_{fc}(r)  \; .  
\end{eqnarray}

The spin--overlaps defined in eq.~(\ref{e29}) are now found to be
\begin{equation}
\begin{array}{lclcl}
C_{0, s_c-1/2}^{1 2} & = & C_{0, s_c-1/2}^{1 3} & = & 
                     - \sqrt{\frac{s_c}{2 s_c +1}}  \; ,  \\
C_{0, s_c+1/2}^{1 2} & = & - C_{0, s_c+1/2}^{1 3} & = & 
                       \sqrt{\frac{s_c+1}{2 s_c +1}} \; , \\
C_{s_c-1/2, s_c-1/2}^{2 3} & = & C_{s_c+1/2, s_c+1/2}^{2 3} & = & 
                     - \frac{1}{2 s_c +1}  \; , \\
C_{s_c-1/2, s_c+1/2}^{2 3} & = & - C_{s_c+1/2, s_c-1/2}^{2 3} & = & 
                       \frac{\sqrt{4 s_c (s_c + 1)}}{2 s_c +1}  \; .
\end{array}
\end{equation}

The Pauli principle requires that the solutions are antisymmetric in a
simultaneous interchange of all the coordinates of the two fermions
labeled 2 and 3. This means that exchange of $\alpha_2$ and $\alpha_3$
and exchange of the order of the couplings in the spin functions must
give a change of sign of the total wave function. Imposing this
constraint the components of the three-body wave function in
eq.~(\ref{e21}) must be related as
\begin{equation} \label{e173}
\phi^{(3)}_{s_c-1/2}=\phi^{(2)}_{s_c-1/2},
\hspace{1cm}
\phi^{(3)}_{s_c+1/2}=-\phi^{(2)}_{s_c+1/2} \; .
\end{equation}

The corresponding Faddeev equations obtained from eq.~(\ref{e27})
after integrating away the spin degrees of freedom are then given by
\begin{eqnarray}
 & \hspace{-5cm} \left[
-\frac{\partial}{\partial \alpha_1^2} -
      (\tilde{\lambda}(\rho)-\rho^2 v_{ff}(\rho \sin{\alpha_1}))
   \right] \phi^{(1)}_0 (\rho, \alpha_1) = -\rho^2 v_{ff}(\rho \sin{\alpha_1})
 \nonumber \\ & \times \left[
\frac{1}{\sin(2\varphi)}
\left( C_{0,s_c-1/2}^{1 2} 
       \int_{|\varphi-\alpha_1|}^{\pi/2 - |\pi/2 - \varphi
-\alpha_1|}  d\alpha_2  \phi_{s_c-1/2}^{(2)}(\rho, \alpha_2) 
\right. \right. \nonumber \\ 
& \left. 
+    C_{0,s_c+1/2}^{1 2} 
       \int_{|\varphi-\alpha_1|}^{\pi/2 - |\pi/2 - \varphi
-\alpha_1|}  d\alpha_2 \phi_{s_c+1/2}^{(2)}(\rho, \alpha_2) \right) 
 \label{e174} \\ & 
+ \frac{1}{\sin(2\tilde\varphi)}
\left( C_{0,s_c-1/2}^{1 2} 
     \int_{|\tilde\varphi-\alpha_1|}^{\pi/2 - |\pi/2 - \tilde\varphi
-\alpha_1|} d\alpha_3 \phi_{s_c-1/2}^{(3)}(\rho, \alpha_3) 
\right. \nonumber \\ 
& \left. \left.
-    C_{0,s_c+1/2}^{1 2} 
     \int_{|\tilde\varphi-\alpha_1|}^{\pi/2 - |\pi/2 - \tilde\varphi
-\alpha_1|} d\alpha_3 \phi_{s_c+1/2}^{(3)}(\rho, \alpha_3) \right)
                                                    \right]  \; , \nonumber
\end{eqnarray}

\begin{eqnarray}
& \hspace{-7.7cm} \left[
-\frac{\partial}{\partial \alpha_2^2} - 
(\tilde{\lambda}(\rho)-\rho^2 v_{fc}^\mp(\rho \sin{\alpha_2}))
           \right] \phi^{(2)}_{s_c \mp 1/2} (\rho, \alpha_2) = 
\nonumber \\ & \hspace{-2.7cm}
-\rho^2 v_{fc}^\mp(\rho \sin{\alpha_2})
\left[ C_{0,s_c \mp 1/2}^{1 2} \frac{1}{\sin(2\varphi)} 
\int_{|\varphi-\alpha_2|}^{\pi/2 - |\pi/2 - \varphi
-\alpha_2|} d\alpha_1 \phi_0^{(1)}(\rho, \alpha_1) + 
\right. \label{e176} \\ & 
\frac{1}{\sin(2\tilde\varphi)} \left( 
C_{s_c \mp 1/2,s_c-1/2}^{2 3} 
\int_{|\tilde\varphi-\alpha_2|}^{\pi/2 - |\pi/2 - \tilde\varphi
-\alpha_2|} d\alpha_3 \phi_{s_c-1/2}^{(3)}(\rho, \alpha_3)
\right.
 \nonumber \\ 
& \left. \left.
+C_{s_c \mp 1/2,s_c+1/2}^{2 3} 
\int_{|\tilde\varphi-\alpha_2|}^{\pi/2 - |\pi/2 - \tilde\varphi
-\alpha_2|} d\alpha_3 \phi_{s_c+1/2}^{(3)}(\rho, \alpha_3) \right)
                                                    \right] \; ,
\nonumber
\end{eqnarray}
where the reduced potentials, $v_{ff} = 2m V_{ff} /\hbar^2, v_{fc}^\mp
= 2m V_{fc}^\mp /\hbar^2$, are defined as in the previous
sections. The two equations with second derivative of $\alpha_3$,
analogous to those of eq.~(\ref{e176}), turn out to be identical to
eq.~(\ref{e176}) due to the constraints in eq.~(\ref{e173}). We
have then three independent Faddeev equations.

The angles $\varphi$ and $\tilde\varphi$ obtained from
eqs.~(\ref{e25}) and (\ref{e24}) are given explicitly by
\begin{equation}
 \varphi =\arctan\left(\frac{M+2m}{M}\right)^{1/2}  \; ,
\end{equation} 
\begin{equation}
 \tilde{\varphi}=\arctan\left(\frac{M(M+2m)}{m^2}\right)^{1/2} \; ,
\end{equation} 
where $M=m_1$ is the mass of the core and $m=m_2=m_3$ the mass of
each of the two fermions.

We shall again use the square-well potentials for the radial shapes of
the two-body potentials between both fermion and core and between the
two fermions, i.e.
\begin{eqnarray}
V_{ff}(r)&=&-V^{(ff)}_0 \Theta(r<R_{ff}) \; ,\\
V_{fc}(r)&=&-V^{(fc)}_0 \Theta(r<R_{fc}) \; .
\end{eqnarray}
The related reduced potentials are then defined as
\begin{eqnarray}
v_{ff}(x)&=&-v^{(ff)}_0\Theta(x<X_{ff}) \; , \\
v_{fc}(x)&=&-v^{(fc)}_0\Theta(x<X_{fc}) \; ,
\end{eqnarray}
and the depth parameters for the two fermion-core relative spin states are 
\begin{equation}\label{e185}
v^{(fc+)}_0(x)= -v^{(fc)}_0(1+ \gamma_s \frac{s_c}{2})  \; , 
\end{equation}
\begin{equation}\label{e194}
v^{(fc-)}_0(x)= -v^{(fc)}_0(1- \frac{\gamma_s (s_c+1)}{2})  \; ,
\end{equation}
where $X_{fc}=R_2\mu_{13}$ and $X_{ff}=R_1\mu_{23}$ are defined as in
section 4.

We shall also in this section restrict ourselves to small or large
distances and omit the lengthy but straightforward calculations and
expressions for the less interesting intermediate distances.

\subsection{Short-distance behavior: $0 \leq \rho \leq {\rm Min}
(X_{ff},X_{fc})$}
The potentials are again constants for all $\alpha$--values and,
as in subsection 4.1, eqs. (\ref{e174}) and (\ref{e176}) have the
general solutions
\begin{equation} \label{e178}
\phi^{(i)}_s(\rho,\alpha_i)=a_{s,i} \sin(2n\alpha_i) \; ,
\end{equation}
where the constants $a_{0,1}$, $a_{s_c\pm 1/2,2}$, and $a_{s_c \pm
1/2,3}$ are found by substitution of eq.~(\ref{e178}) into
eqs. (\ref{e174}) and (\ref{e176}). We obtain the following system of
linear equations for
\begin{eqnarray} \label{e181}
\lefteqn{
 ( \epsilon - v^{(ff)}_0) a_{0,1}=} \label{e182}\\ &
  - v^{(ff)}_0 \left[ d C^{12}_{0, s_c-1/2} (a_{s_c-1/2,2}+a_{s_c-1/2,3})
                      +d C^{12}_{0, s_c+1/2} (a_{s_c+1/2,2}-a_{s_c+1/2,3})
                                                          \right] \; ,
\nonumber
\end{eqnarray}
\begin{eqnarray} \label{e183}
\lefteqn{
 (\epsilon -v^{(fc\mp)}_0) a_{s_c \mp 1/2,2}=}  
                                                 \\ &
 -  v^{(fc\mp)}_0 \left[ d C^{12}_{0, s_c\mp 1/2} a_{0,1}+
          \tilde{d} (C^{23}_{s_c\mp 1/2, s_c-1/2} a_{s_c-1/2,3}+
                     C^{23}_{s_c\mp 1/2, s_c+1/2} a_{s_c+1/2,3}) \right] \; ,
              \nonumber
\end{eqnarray}
where we in analogy with eq.~(\ref{e120}) define
\begin{eqnarray}
\tilde{\lambda}(\rho)=4 n^2 - \rho^2 \epsilon \; ,
  \hspace{1cm}
d= - \frac{1}{n} \frac{\sin(2n \varphi)}{\sin(2\varphi)} \; ,
  \hspace{1cm}
\tilde{d}= - \frac{1}{n} \frac{\sin(2n \tilde{\varphi})}{\sin(2\tilde{\varphi})}
  \; .\label{e184}
\end{eqnarray}

The antisymmetry expressed in eq.~(\ref{e173}) relate the unkown
constants by
\begin{eqnarray} \label{e175}
a_{s_c-1/2,3}=a_{s_c-1/2,2} \; \; ,
a_{s_c+1/2,3}=-a_{s_c+1/2,2}  \; ,
\end{eqnarray}
which by use in eqs.~(\ref{e181}) and (\ref{e183}) reduces the number
of equations and the number of unknown constants to three.

The solutions for $\epsilon$ or $\tilde{\lambda}(\rho)$ are now
obtained by demanding non-vanishing solutions for the constants
$a_{s,i}$. This amounts effectively to three linear equations and the
corresponding determinant must vanish. Thus $\epsilon$ is
obtained from
\begin{equation}
\left|
\begin{array}{ccc}
  \epsilon-v^{(ff)}_0  &  2v^{(ff)}_0 d C^{12}_{0,s_c-1/2}  
                   &  2v^{(ff)}_0 d C^{12}_{0,s_c+1/2} \\ 
  v^{(fc-)}_0 d C^{12}_{0,s_c-1/2}  &  \epsilon - v^{(fc-)}_0 (1-\tilde{d}
C^{23}_{s_c-1/2,s_c-1/2})  & - v^{(fc-)}_0 \tilde{d} C^{23}_{s_c-1/2,s_c+1/2}
\\
  v^{(fc+)}_0 d C^{12}_{0,s_c+1/2}  & - v^{(fc+)}_0\tilde{d} C^{23}_{s_c-1/2,s_c+1/2}
&\epsilon - v^{(fc+)}_0 (1+\tilde{d} C^{23}_{s_c-1/2,s_c-1/2})

\end{array}
\right| = 0 \; .
\label{e186}
\end{equation}
This means that there are at most three $\lambda$-solutions for each $n$;
sometimes less than three, since some of them can be the trivial
spurious solutions characterized by $\epsilon=0$.

When the determinant vanishes for $\epsilon=0$, we find that these spurious
solutions must satisfy one or both of the following two conditions:
\begin{equation}
n-\frac{\sin(4n\varphi)}{\sin(4\varphi)}=0  \; ,
\label{spur1}
\end{equation}
\begin{equation}
n^2+\frac{\sin(4n\varphi)}{\sin(4\varphi)} n - 2
\frac{\sin^2(2n\varphi)}{\sin^2(2\varphi)} = 0 \; .
\label{spur2}
\end{equation}

For $n=1$ ($d=\tilde{d}=-1$) these two equations are both satisfied, 
meaning that two of
the solutions are spurious and only one antisymmetric solution exists.
Solving the determinant (\ref{e186}) the antisymmetric solution is found 
to be
\begin{equation}
\tilde{\lambda}(\rho)=\lambda(\rho)+4=4-\rho^2 
\left( v^{(ff)}_0 + v^{(fc-)}_0 \frac{2 s_c}{2s_c+1} + v^{(fc+)}_0 
                         \frac{2s_c+2}{2s_c+1} \right),
\end{equation}
which reduces to eq.~(\ref{e117}) for identical and spin independent
potentials.

For $n=2$, where $2d^2= 1 - \tilde{d}$, only the condition in
eq.~(\ref{spur2}) is satisfied, and two non-spurious antisymmetric
solutions appear. They are the solutions of the second order equation
\begin{eqnarray}
\epsilon^2&-& \epsilon \left( v^{(ff)}_0 + v^{(fc+)}_0 + v^{(fc-)}_0  
  + \frac{\tilde{d}}{2s_c+1} (v^{(fc-)}_0 - v^{(fc+)}_0) \right) \\ 
  &+ &
(1+\tilde{d}) \left( v^{(ff)}_0 v^{(fc-)}_0 \frac{s_c+1}{2s_c+1}
+ v^{(ff)}_0 v^{(fc+)}_0 \frac{s_c}{2s_c+1} + v^{(fc+)}_0 v^{(fc-)}_0 
 (1-\tilde{d})   \right) = 0 \;  ,\nonumber
\end{eqnarray}
which combined with eq.(\ref{e184}) results in two solutions for
$\tilde{\lambda}(\rho)$.

For $n \geq 3$ none of the conditions (\ref{spur1}) and (\ref{spur2})
are satisfied, and three non-spurious antisymmetric solutions are
found.  For very big values of $n$ both $d$ and $\tilde{d}$ approach
zero, and the three solutions of the determinant converges towards
$\epsilon=v^{(ff)}_0$, $\epsilon=v^{(fc-)}_0$ and $\epsilon=v^{(fc+)}_0$.

\subsection{Large-distance behavior}
As in eqs.(\ref{e57}) and (\ref{e111}), we define
\begin{eqnarray}
\alpha_0^{(ff)}&=&\arcsin(X_{ff}/\rho) \; ,\\ 
\alpha_0^{(fc)}&=&\arcsin(X_{fc}/\rho)
\end{eqnarray}
such that the potential is non-vanishing only when the corresponding
$\alpha \leq \alpha_0$.

Since $\alpha_0^{(ff)}$ and $\alpha_0^{(fc)}$ approach zero for increasing
$\rho$, we can define large distances by
\begin{eqnarray}
\alpha_0^{(ff)} \leq \alpha_0^{(fc)} \leq |\varphi - \alpha_0^{(ff)}|
\leq |\varphi - \alpha_0^{(fc)}| \label{e188} \; , \\
\alpha_0^{(ff)} \leq \alpha_0^{(fc)} \leq |\tilde \varphi - \alpha_0^{(ff)}|
\leq |\tilde \varphi - \alpha_0^{(fc)}| \; ,
\label{e190}
\end{eqnarray}
where we assumed that $X_{ff} < X_{fc}$.

\begin{figure}[t]
\epsfxsize=12cm
\epsfysize=7cm
\epsfbox[-30 -400 520 -50]{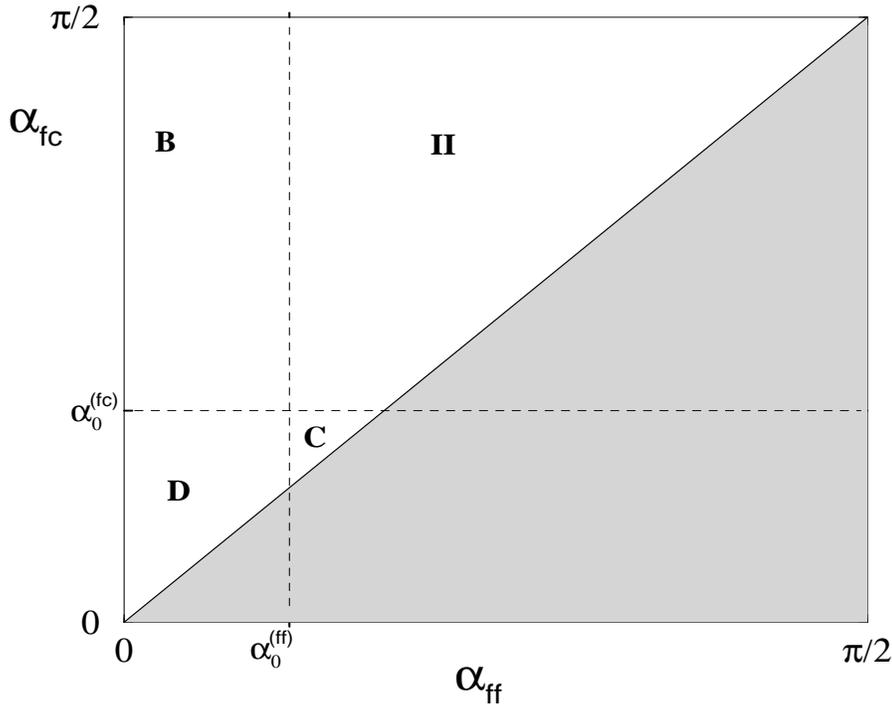}
\vspace{3.5cm}
\caption{\protect\small
The regions, defined in subsection 5.2, in
$(\alpha_{fc},\alpha_{ff})-$space arising for large distances for two
identical spin-1/2 particles. The shaded area does not enter in the
present computation.
}
\end{figure}

In fig.~4 we plot $\alpha_{fc}=\alpha_2$ versus $\alpha_{ff}=\alpha_1$
where only the  region, $\alpha_{ff}<\alpha_{fc}$, is relevant for our
calculation. In the plot we can distinguish four different zones:
\begin{eqnarray}
II  & :& \alpha_{fc}>\alpha_0^{(fc)} \mbox{\ and\ } \alpha_{ff}>\alpha_0^{(ff)}
          \Longrightarrow V_{fc}(\rho \sin\alpha_{fc})=0 \mbox{\ and\ } 
     V_{ff}(\rho \sin\alpha_{ff})=0 
               \nonumber \\
B  & :& \alpha_{fc}>\alpha_0^{(fc)} \mbox{\ and\ } \alpha_{ff}<\alpha_0^{(ff)}
          \Longrightarrow V_{fc}(\rho \sin\alpha_{fc})=0 \mbox{\ and\ } 
  V_{ff}(\rho \sin\alpha_{ff})\neq 0
               \nonumber \\
C & :& \alpha_{fc}<\alpha_0^{(fc)} \mbox{\ and\ } \alpha_{ff}>\alpha_0^{(ff)}
          \Longrightarrow V_{fc}(\rho \sin\alpha_{fc})\neq 0 \mbox{\ and\ } 
   V_{ff}(\rho \sin\alpha_{ff})=0
               \nonumber \\
D  & :& \alpha_{fc}<\alpha_0^{(fc)} \mbox{\ and\ } \alpha_{ff}<\alpha_0^{(ff)}
          \Longrightarrow V_{fc}(\rho \sin\alpha_{fc})\neq 0 \mbox{\ and\ } 
  V_{ff}(\rho \sin\alpha_{ff})\neq 0
               \nonumber 
\end{eqnarray}
Since $\alpha_0^{(ff)}$ and $\alpha_0^{(fc)}$ are close to zero, regions
$B$, $C$, and $D$ are very small. Due to the inequalities in
eqs.(\ref{e188}) and (\ref{e190}), the integrations appearing in
the Faddeev equations involve only the functions in the region where
the potentials vanish.

The wave functions must vanish at $\alpha=0$ and $\alpha=\pi/2$ and
the solutions in region II, where all potentials are zero, are
therefore of the form
\begin{equation}\label{e187}
 \phi^{(i)}_s(\alpha_i) = b_{s,i} \sin \left((\alpha_i - \pi/2)
  \sqrt{\tilde\lambda} \right)  \; .
\end{equation}
In the other regions (B,C and D) the same form of the wave function is
a solution when the corresponding potentials vanish, i.e. for
$\alpha_{i}>\alpha_0^{(i)}$. When $\alpha_{i}<\alpha_0^{(i)}$ in these
regions, we have solutions of the form
\begin{equation}\label{e189}
  \phi^{(i)}_s(\alpha_i) = c_{s,i} \sin (\alpha_i \kappa_s)
 + d_{s,i} \sin \left(\alpha_i  \sqrt{\tilde\lambda} \right)  \;  ,
\end{equation}
where the first term is the solution to the homogeneous equation and
consequently the different $\kappa_s$-values are given by
\begin{equation}\label{e191}
 \kappa_0=\sqrt{v^{(ff)}_0 \rho^2 +\tilde{\lambda}(\rho)} \; , \; \; 
 \kappa_{s_c \pm 1/2}=\sqrt{v^{(fc\pm)}_0 \rho^2 +\tilde{\lambda}(\rho)} \; .
\end{equation}
The constants $c_{s,i}$ are therefore completely free, since the
functions to be integrated in eqs.(\ref{e174}) and (\ref{e176}) only
involve the functions in eq.(\ref{e187}), i.e. values of
$\alpha_{i}>\alpha_0^{(i)}$ for all three components, see fig.~4. The
constants $b_{s,i}$ and $d_{s,i}$ are linearly related through the
Faddeev equations.

The detailed solutions in the different regions are given in appendix
F. The eigenvalue equation again takes the form of a vanishing
determinant where the matrix elements are given in eq.(F\ref{ef19}).

In the limit of very large distances we can for finite
$\tilde{\lambda}-$values use an expansion to lowest order in $1/\rho$
as given in appendix F, where also the case of diverging
$\tilde{\lambda}-$values are considered. The linear set of equations
in eq.(F\ref{ef15})  then reduces to
\begin{eqnarray}
A_1 \left(\sqrt{\tilde{\lambda}} \cos(\frac{\pi}{2} \sqrt{\tilde{\lambda}})
\frac{a_{scat}^{(ff)}}{\rho} \mu_{23} + \sin(\frac{\pi}{2} 
 \sqrt{\tilde{\lambda}}) \right)
= -a_{0} \sqrt{\tilde{\lambda}} \frac{a_{scat}^{(ff)}}{\rho} \mu_{23} \\
A_2 \left(\sqrt{\tilde{\lambda}} \cos(\frac{\pi}{2} \sqrt{\tilde{\lambda}})
\frac{a_{scat}^{(fc-)}}{\rho} \mu_{12} + \sin(\frac{\pi}{2} 
 \sqrt{\tilde{\lambda}}) \right)
 = -a_{s_c-1/2} \sqrt{\tilde{\lambda}}  \frac{a_{scat}^{(fc-)}}{\rho} \mu_{12} \\
A_3 \left(\sqrt{\tilde{\lambda}} \cos(\frac{\pi}{2} \sqrt{\tilde{\lambda}})
\frac{a_{scat}^{(fc+)}}{\rho} \mu_{12} +  \sin(\frac{\pi}{2} 
 \sqrt{\tilde{\lambda}})\right)
=   -a_{s_c+1/2} \sqrt{\tilde{\lambda}} \frac{a_{scat}^{(fc+)}}{\rho} \mu_{12}
  \; ,
\end{eqnarray}
where $\mu_{12}$ and $\mu_{23}$ are constants defined in appendix A, and
$a_0, a_{s_c \pm 1/2}$ depend linearly on the coefficients $A_i$ as
defined in appendix F. The eigenvalues for $\rho \rightarrow \infty$
therefore for finite scattering lengths approach solutions to
$\sin(\frac{\pi}{2}\sqrt{\tilde\lambda}) = 0$, i.e.\ the
hyperspherical spectrum of $\tilde\lambda = 4n^2$.

The solutions to the next order in $1/\rho$ only exist when the
corresponding determinant is zero which then defines the behavior of
$\tilde{\lambda}(\rho)$ for large values of $\rho$.  We have
previously assumed that $\alpha_0^{(ff)} < \alpha_0^{(fc)}$ and both
quantities are close to zero in the large-distance limit. If we
further assume that $\alpha_0^{(ff)}=0$ or equivalently $X_{ff}=0$, only
regions II and C survive in fig.4. Then $A_1=0$, $|a_{scat}^{(ff)}|/\rho
\approx 0$ (for finite $a_{scat}^{(ff)}$ and large $\rho$) and the
fermion-fermion interaction completely disappears. The determinant is
reduced to a 2 by 2 determinant, that gives the following expression
for $\tilde{\lambda}(\rho)$
\begin{eqnarray} \label{e193}
&&
\frac{\rho^2}{a_{scat}^{(fc+)}a_{scat}^{(fc-)}}
\sin^2\left(\frac{\pi}{2} \sqrt{\tilde{\lambda}}\right)
+\frac{1}{2} \sqrt{\tilde{\lambda}} \sin(\pi \sqrt{\tilde{\lambda}})
\mu_{12} \left( \frac{\rho}{a_{scat}^{(fc+)}} + \frac{\rho}{a_{scat}^{(fc-)}} 
   \right)    \nonumber     \\ && 
+\frac{2}{2s_c+1}\left(\frac{\rho}{a_{scat}^{(fc-)}}-
 \frac{\rho}{a_{scat}^{(fc+)}}\right) \tilde{f} 
\sin\left(\frac{\pi}{2} \sqrt{\tilde{\lambda}}\right) \\ &&
+ \tilde{\lambda} \cos^2\left(\frac{\pi}{2} \sqrt{\tilde{\lambda}}\right)
\mu_{12}^2  - 4 \mu_{12}^2 f^2 = 0   \nonumber  \; ,
\end{eqnarray}
where $f$ and $\tilde{f}$ are defined in eq.(F\ref{ef4}).

In the limit where both scattering lengths are large compared to
$\rho$, i.e. $|a_{scat}^{(fc\pm)}| \gg \rho$, we arrive again at
eq.(\ref{e161}). The occurrence conditions and properties of the
Efimov states can be derived from eq.(\ref{e193}). The details are
discussed in \cite{fed95}.

The behavior of $P_{nn^\prime}$ and $Q_{nn^\prime}$ at large distance
is qualitatively the same as for three spinless particles. This result
is again, along with the various solutions described in this
subsection, valid in general for all short-range potentials. In
particular, also the Efimov conditions remain the same for such
potentials.

\section{Numerical illustrations and generalizations}
The analytical solutions in the previous sections are derived for
square-well potentials. The results are similar for other short-range
potentials. Illustrations by use of smoother gaussian potential are
therefore appropriate and we shall first show numerically calculated
angular eigenvalue spectra for different symmetries of the three-body
system. Then we shall give a survey of the analytical results for
different distances and show how the results can be used for arbitrary
short-range potentials. The simplest case of intrinsic spins of the
particles were also considered in the previous sections. Other cases
with more complicated spin structures can be worked out. It is
straightforward, but results easily in rather extended formulae. We
shall here indicate how to proceed in the case of two different
spin-1/2 particles.

\begin{figure}[t]
\epsfxsize=12cm
\epsfysize=7cm
\epsfbox[-30 -400 520 -50]{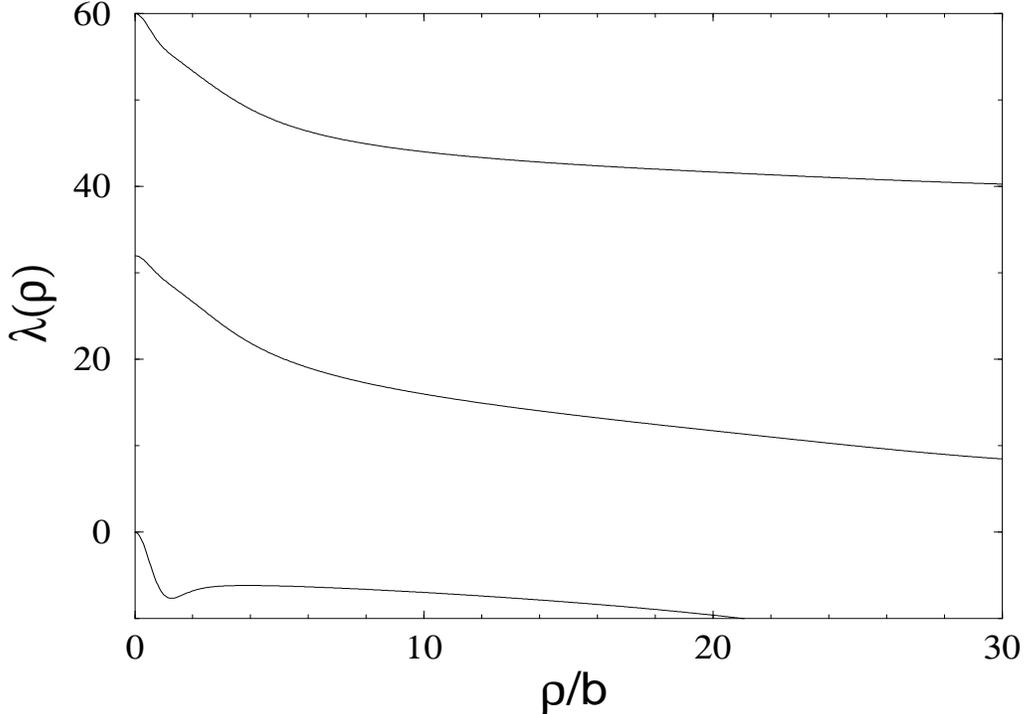}
\vspace{3.5cm}
\caption{\protect\small
The lowest angular eigenvalues $\lambda =
\tilde\lambda - 4$, when all orbital angular momenta are zero, are
shown as function of $\rho/b$ for the case of three identical bosons
for a gaussian potential, $\bar{V}_0\hbar^2/mb^2$~exp(-$(r/b)^2$),
where $m$ is the mass of the bosons. The actual value of
$\bar{V_0}=-3.08$ corresponds to a bound two-body state.
}
\end{figure}

\subsection{Numerical illustrations}
The angular eigenvalues, still only s-states, for the symmetric case
of three identical bosons are shown in fig. 5 as function of $\rho /b$
for a gaussian potential. The parameters of the potential correspond
to a two-body bound state. The lowest eigenvalue consequently bends
over and diverges parabolically as described in eq.(\ref{e90}). The
higher lying levels then come down and the hyperspherical spectrum
seen at $\rho=0$ is approached at large distances. The values of
$\lambda = 4n^2-4$ at $\rho=0$ are 0, 32, 60 and the spurious state
starting from $\lambda=12$ corresponding to $n=2$ in eq.(\ref{e53}) is
omitted. Due to the symmetry requirement only one state appears for
each value of n.

\begin{figure}[t]
\epsfxsize=12cm
\epsfysize=7cm
\epsfbox[-30 -400 520 -50]{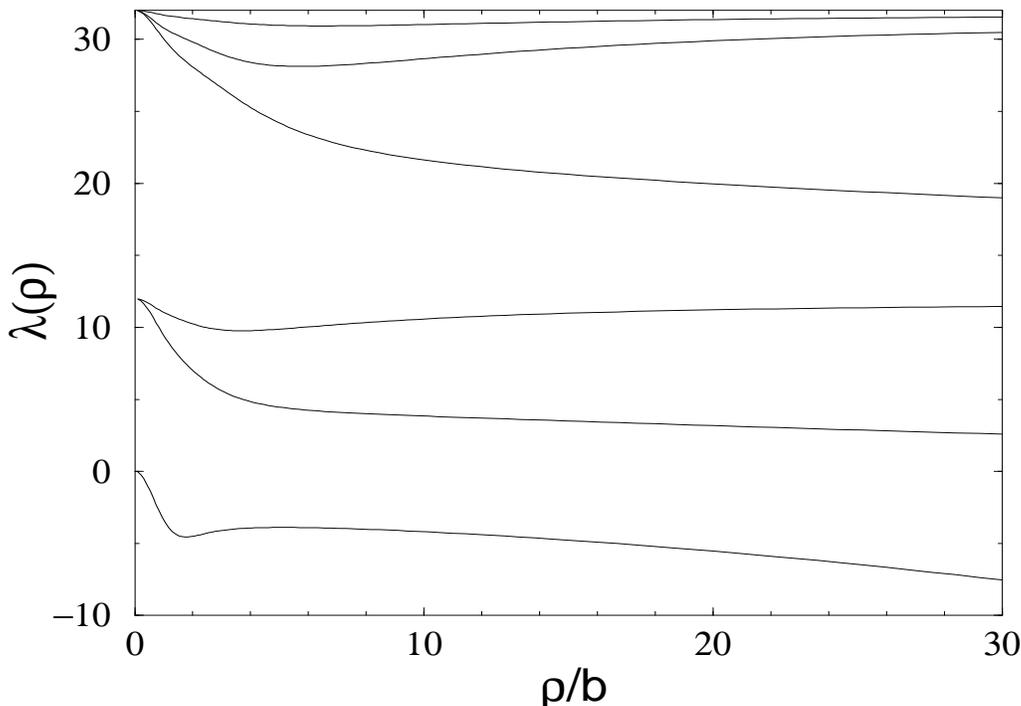}
\vspace{3.5cm}
\caption{\protect\small
The lowest angular eigenvalues $\lambda =
\tilde\lambda - 4$, when all orbital angular momenta are zero, are
shown as function of $\rho/b$ for three different spinless particles
for gaussian potentials,
$\bar{V}_0^{(i)}\hbar^2/m_ib^2$~exp(-$(r/b)^2$). The actual values
$\bar{V}_0^{(i)}=-1.28, -1.54, -3.08$ and $m_i=m_1, 2m_1, 3m_1$,
respectively for $i=1,2,3$, correspond to a bound two-body state in
the subsystem labeled i=1 and no bound states in the other subsystems.
}
\end{figure}

In fig. 6 we show the angular eigenvalues as function of $\rho /b$,
again only s-states, for the asymmetric case of three different
spinless particles interacting via gaussian potentials. Only one of
the potentials has a bound two-body state and consequently the lowest
eigenvalue bends over and diverges parabolically as described in
eq.(\ref{e162}). Again the higher lying levels come down and the
spectrum at $\rho=0$ emerges also in this case at large distances. The
spurious level corresponding to n=2 is still omitted, but now two
other levels appear at $\lambda=12$ corresponding to (spatially)
non-symmetric configurations. Also for n=3 at $\lambda=32$ two more
levels corresponding to non-symmetric states appear in addition to the
totally symmetric state. For all values of $n \ge 3$ the structure of
one symmetric and two asymmetric states remains unchanged.

In fig. 7 we show analogous numerical results for two identical
spin-1/2 fermions plus one third particle (core) of spin
$s_c=3/2$. Now one additional degree of freedom appears. It is related
to the two different fermion-core spin couplings. However,
simultaneously the two fermions are restricted to totally
antisymmetric two-body states. This clearly removes some of the
otherwise possible states. Since the hyperspherical spectrum always
both is the starting point at $\rho=0$ and the asymptotic limit for
large $\rho$, the combined result is a spectrum very similar to that
of fig. 6 corresponding to three different particles without spin
degrees of freedom. If the core had been spinless, the fermion-core
spin coupling had been unique and one state less would have appeared
for every value of $n \ge 2$.

\begin{figure}[t]
\epsfxsize=12cm
\epsfysize=7cm
\epsfbox[-30 -400 520 -50]{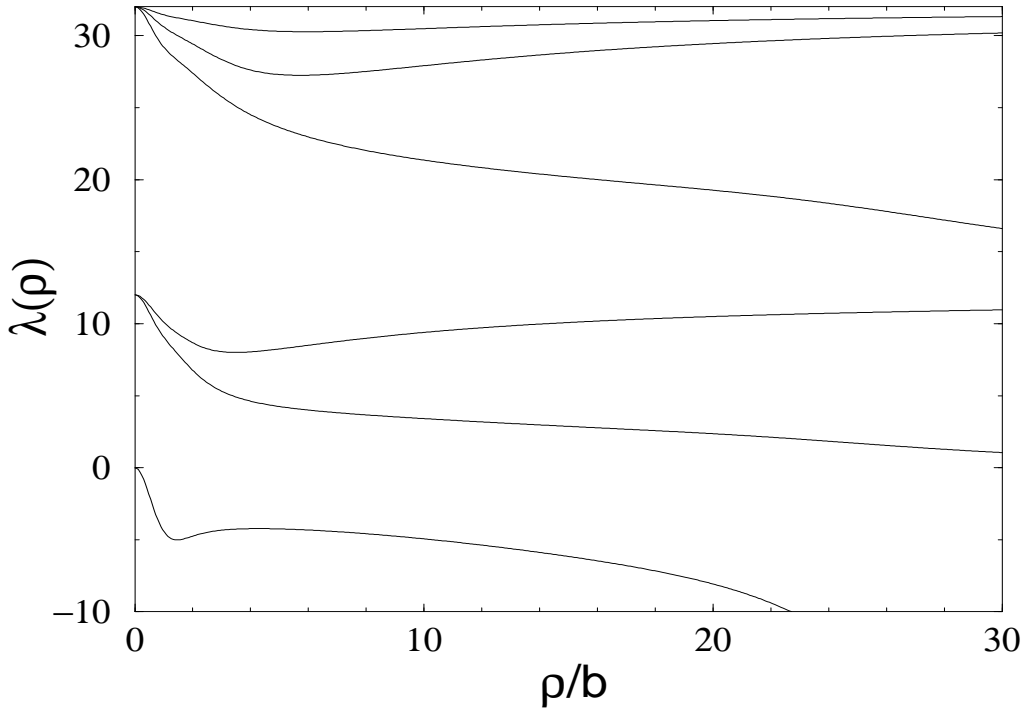}
\vspace{3.5cm}
\caption{\protect\small
The lowest angular eigenvalues $\lambda =
\tilde\lambda - 4$, when all orbital angular momenta are zero, are
shown as function of $\rho/b$ for gaussian potentials,
$\bar{V}_0^{(i)}\hbar^2/m_ib^2$~exp(-$(r/b)^2$), for a system of two
identical spin-1/2 fermions plus a third particle of spin
$s_c=3/2$. The actual values $\bar{V}_0^{(i)}=-6.16, -0.99, -1.09 $
and $m_i=2m_2, m_2, m_2$, respectively for $i=1,2,3$, correspond to
the fermion-fermion interaction, the fermion-core interaction for
relative spin of 2 and the fermion-core interaction for relative spin
of 1, respectively for $i=1,2$ and 3.
}
\end{figure}

\subsection{Survey and generalization of the solutions}
Various analytical solutions apply to different regions in the
two-dimensional $(\rho , \alpha)$ coordinate space. A survey is shown
in fig. 8 where the regions labeled I,A,B,C and II refer to the
division in the subsections of the symmetric case. (The more general
asymmetric case requires more divisions at intermediate
distances. However, the picture basically remains unchanged.) The
simplest cases of small and large distances (I and II) actually
constitute a very large fraction of the total coordinate space. These
solutions are also appropriate for general short-range potentials
where the approximate validity results from perturbative treatments.

At large distances, where $\alpha_0$ and $\kappa$ can be expanded to
first order in $1/\rho$, we obtain considerable further
simplification. The eigenvalue equation can in this limit be expressed
entirely in terms of the scattering length of the potential.  In
fig.~9 we show the lowest angular eigenvalue for a square-well
potential as function of $\rho$. We compare with the approximate
solution at very large distances given in eq.(\ref{e104}). The
$1/\rho$ asymptotic convergence towards the limit is clearly
seen. This approximation is normally too inaccurate for the
interesting shorter distances.

A numerical procedure for general short-range potentials now suggest
itself. First we construct the equivalent square-well potential with
the same scattering length and effective range as the potential in
question.  The solutions are then obtained analytically in
$\alpha-$space as described in this paper. They can be anticipated to
be very accurate solutions except in smaller regions at intermediate
distances, where direct numerical integration interpolating between
the available analytical solutions then must be used. The gain in
accuracy and speed is substantial.

An example is shown in fig.~9 where we used an attractive gaussian
potential of range $b$ and strength $-2.19\hbar^2/mb^2$ with a
scattering length, $a_{scat}=5b$, and an effective range,
$R_{eff}=1.64b$, where b is the range of the potential. The
corresponding radius and depth of the square-well potential is then
$R_0=1.47b$ and $V_0=-0.93\hbar^2/mb^2$, respectively.

\begin{figure}[t]
\epsfxsize=12cm
\epsfysize=7cm
\epsfbox[-30 -400 520 -50]{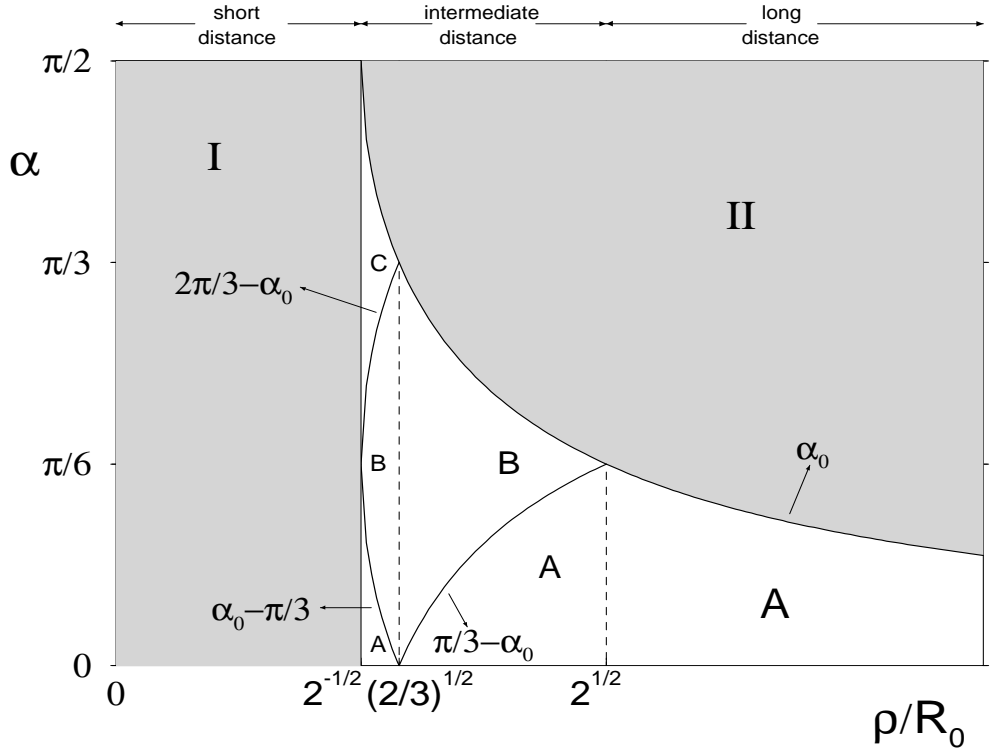}
\vspace{3.5cm}
\caption{\protect\small
The regions in $\alpha$-space as function of
$\rho/R_0$ for a square-well potential of radius $R_0$ for three
identical bosons. The shaded areas indicate the small and
large-distance regions I and II, where the wave functions are
particularly simple.
}
\end{figure}

The curves for the two potentials are indistinguishable at large
distances down to $\rho = R_0\sqrt{2} \approx 2b$. Then the square-well result is lowest until $\rho \approx b$, where the curves cross
and the gaussian result stays as the lowest all the way down to zero.
The largest deviation is less than about 0.75 compared to the minimum
value of about -6. This relatively small region, $0.3 \le \rho /b \le
2$, at intermediate distances, corresponds to the surface region for
the two-body potential. The perturbative short-distance solution for
the gaussian potential is accurate up to about $\rho \approx
0.3b$. The large-distance behavior describing the approach towards the
asymptotic limit is only a reasonable approximation at very large
distances.

The exact square-well solution is almost quantitatively a good
approximation. The two energies would be close due to the similarities
and the fact that a larger potential at smaller distance is
compensated by a smaller potential at larger distance. The square well
tends to confine the wave function in a somewhat more narrow region
around the minimum.

The (smaller) differences between the two potentials occur in an
essential region, where a substantial part of the attractive pocket of
the effective three-body potential is contained. High accuracy for an
arbitrary potential therefore requires a treatment better than that
corresponding to the square-well solution. In practise the most
efficient procedure is numerical integration starting with the
small-distance perturbative solution. Reaching large distances
corresponding to $\rho \le R_0\sqrt{2}$ the exact square-well
solution, which then is particularly simple, can be used to a very high
accuracy.

\subsection{Two different spin-1/2 particles}
We consider a system consisting of two non-identical spin-1/2
particles and a third particle of arbitrary spin $s_c$. The procedure
is then the same as in section 5, but now the spin state
$\chi_1^{(1)}$ is allowed and the antisymmetrization constraint
eq.(\ref{e173}) is not required. The three Faddeev equations can then
be written as a system of six equations, where the angles
$\varphi_{1}$, $\varphi_{2}$, and $\varphi_{3}$ all are
different. Since the two fermions can couple to spin 1, the total spin
of the three-body system can be $J=s_c,s_c\pm 1$. In case when
$J=s_c\pm 1$, the spin states $\chi_0^{(1)}$, $\chi^{(2)}_{s_c\mp
1/2}$, and $\chi^{(3)}_{s_c \mp 1/2}$ are not possible and the total
wave function is symmetric under exchange of the two fermions, and the
number of Faddeev equations is again reduced to three.

\begin{figure}[t]
\epsfxsize=12cm
\epsfysize=7cm
\epsfbox[-30 -400 520 -50]{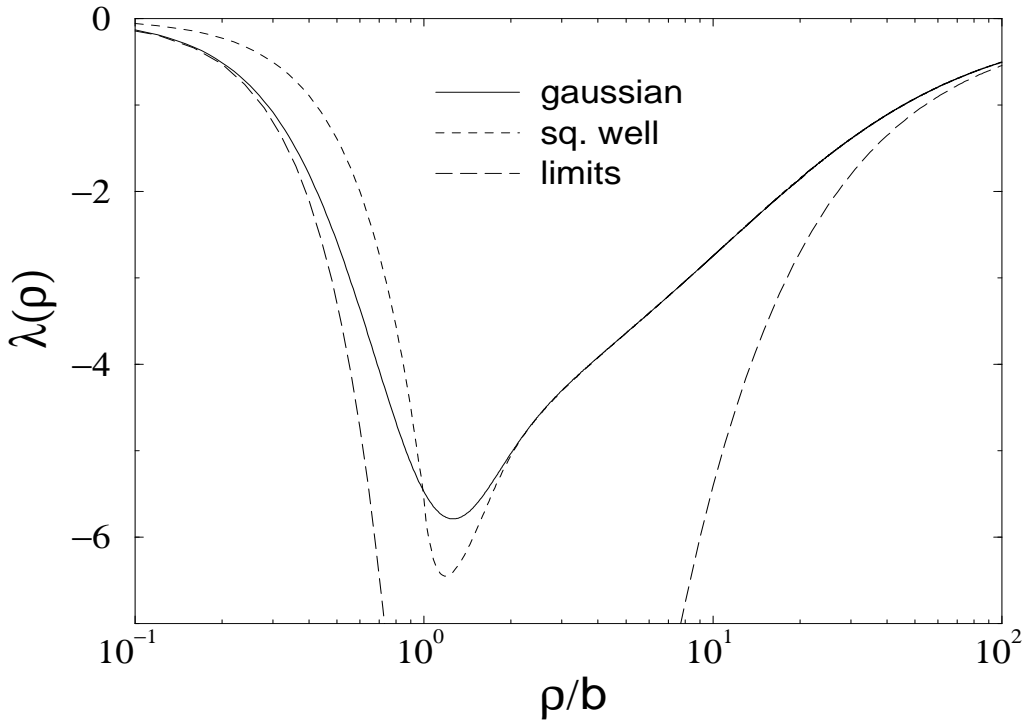}
\vspace{3.5cm}
\caption{\protect\small
The lowest angular eigenvalue $\lambda =
\tilde\lambda - 4$ as function of $\rho/b$ for a system of three
identical bosons. The interactions correspond to a gaussian potential
(solid curve), $\bar{V_0}\hbar^2/mb^2$~exp(-$(r/b)^2$),
$\bar{V_0}=-2.19$, and a square-well potential (short-dashed curve),
radius $R_0=1.47b$ and depth $-0.93\hbar^2/mb^2$, with the same
scattering length $a_{scat}=5b$, and effective range
$R_{eff}=1.64b$. The perturbative solution for the gaussian potential
at small distance (eq.~(\ref{e53})) and the limiting behavior at large
distance (eq.~(\ref{e104})) are also shown (long-dashed curves).
}
\end{figure}

At short-distances the general solutions are given by eq.(\ref{e178}).
They give a 6 by 6 determinant when $J=s_c$ and a 3 by 3 determinant
when $J=s_c \pm 1$. Making these determinants equal to zero and
solving the equation for $\epsilon=0$ one can extract the spurious
solutions and the non-spurious ones for $\epsilon\neq 0$.

At large distances the procedure is also analogous to the one shown in
the subsection 5.2. The only difference is that now we have
$\alpha_0^{(f_1 f_2)}$, $\alpha_0^{(f_1 c)}$, and $\alpha_0^{(f_2
c)}$, and the number of regions we have to consider (similar to
regions II, A, B, and C in subsection 5.2) is much larger. Technically
the problem is more cumbersome, but conceptually it is identical.

\section{Summary and conclusions}

The new method to solve the Faddeev equations in coordinate space has
several advantages especially in the long-distance description. The
reason is the access to the asymptotic analytic properties made
possible by the two-step sequential procedure. The generalized angular
equations are first solved for each average distance (radial
coordinate) between the three particles. The periodic behavior or the
finite intervals characteristic for angular variables provide discrete
quantized solutions which subsequently are used as a complete basis
set in an expansion of the total wave function. The expansion
coefficients (as well as the basis) are functions of the radial
coordinate. They are in the next step determined together with the
total energy from the coupled set of radial equations. We have first
formulated this general procedure for s-states with the inclusion of
intrinsic spins for each of the three particles. For spin-independent
interactions we recover the previous general s-state Faddeev
equations.

We study the symmetric case described by three equal Faddeev
components. We assume spin independent interactions and arrive at one
integro--differential angular Faddeev equation for s-states. This
equation is solved analytically for a square-well potential. The
angular wave functions are especially simple at small (one sine
function) and large distances (one or two sine functions). These
solutions are, with an appropriate interpretation, more general than
their origin as square-well solutions seems to suggest. At small
distance, they are the first order perturbative solutions for an
arbitrary potential with a value at the center equal to the
square-well depth. At sufficiently large distances where the general
short-range potential has vanished the solution obviously is identical
to the square-well solutions outside its radius. We show explicitly
that the coupled set of radial equations decouple at large distances.

At intermediate distances the solutions are linear combinations of up
to four sine functions of varying arguments. The corresponding
eigenvalues are explicitly given at small distance where they vary
parabolically with distance starting from the hyperspherical spectrum.
At intermediate distances the eigenvalues are solutions to cubic
equations and a simple trancendental equation involving trigonometric
functions arises at large distances.

The length scale is here defined by the radius of the square-well
potential and the large-distance solution is valid when the average
radius exceeds $R_0 \sqrt{2}$. Expansion of the eigenvalue equation to
lowest order in the inverse distance brings additional simplification,
but introduces also the two-body scattering length as another length
parameter. The Efimov condition leading to infinitly many bound
three-body states is then obtained and seen to be independent of the
short-range potential.

The asymmetric case, still s-states and spin independent interactions,
is described by three Faddeev equations. We can again solve the
angular part analytically for square-well potentials acting betweeen
each pair of particles. Each Faddeev component of the wave functions
is at small and large distances of the same form as in the symmetric
case. These solutions are again appropriately interpreted solutions to
general potentials either in perturbation theory or asymptotically at
large distances. The corresponding eigenvalues are in this case
obtained either as solutions to a cubic equation or to a trancendental
equation involving trigonometric functions. At intermediate distances
the solutions can still be found analytically to be combinations of
exponentials and trigonometric functions. However, the number of terms
is now substantial and we abstain from writing down these soulutions.

Expanding the eigenvalue equation to first order in the inverse
distance again brings additional simplifications and introduces the
three (different) scattering lengths. The Efimov conditions leading to
infinitly many bound three-body states can then be seen to exist when
at least two scattering lengths are infinitly large. Asymptotic
large-distance behavior of the effective radial potentials are derived
to first order in the inverse distance.

The case of two identical spin-1/2 fermions and one additional core
particle of arbitrary spin can of course also be described by three
Faddeev equations. We still include only s-states and assume a
spin-spin type of interaction between fermion and core
particles. Elimination of the spin degrees of freedom results in three
angular Faddeev equations where the reduction from six to three
equations arise due to the requirement of an antisymmetric wave
function. Again these equations are solved for radial square-well
potentials both for small and large distances. Expansion to first
order in the inverse distance expresses the angular eigenvalue
equation in terms of three (different) scattering lengths. The Efimov
conditions are described and discussed.

We finally considered qualitatively the case of two different spin-1/2
particles interacting mutually and with a third particle. The original
three Faddeev equations results now, still only including s-states, in
sixcoupled angular equations after elimination of the spin degrees
of freedom. Small and large-distance behavior can be studied
analogously for radial square-well potentials. The systems considered
here containing spin-1/2 particles are chosen as the simplest examples
of three interacting particles with an intrinsic structure showing up
in the form of spin degrees of freedom. Other similar examples can be
worked out in the same way.

The angular solutions obtained with square-well potentials have
properties characteristic for short-range potentials. To calculate the
total solution we must in addition solve the coupled set of radial
differential equations. This is a rather simple numerical problem. The
coupled set of radial equations decouple at large distances. Only
s-waves are included in the discussion, but they are usually the most
interesting components in the wave functions. For bound states only
s-waves extend far beyond the radii of the short-range
potentials. They decouple from higher angular momentum components at
large distances and often constitute the dominating part of the total
wave function.

The square-well solutions are intrinsically interesting. Furthermore,
they approach at large distance the solutions for arbitrary
short-range potentials. The exact solutions are valid down to
distances close to the radius of the square-well potentials. The
solutions for general potentials can now be found by adjusting depth
and radius of a square-well potential to obtain the same scattering
length and effective range. The large-distance solutions obtained as
described in this paper are then correct down to $\sqrt{2}$ times the
radius of the square-well potential, i.e. at much smaller distances
than given by the large-distance expansion for general
potentials. Therefore these results can be used to make the numerical
computations both substantially faster and more accurate.

In conclusion, we have discussed a method and derived solutions
providing basic insight into the Faddeev equations. Practical
numerical improvements are suggested for arbitrary short-range
potentials. \\

{\bf Acknowledgments} One of us (DVF) acknowledges the support from
INFN, Trento, Italy. E.G acknowledges support from the European Union
through the Human Capital and Mobility program contract
nr. ERBCHBGCT930320.

\bigskip
\noindent
{\Large \bf Appendix A:  Coordinates}
\setcounter{equation}{0}
\\
\noindent
We consider a system of three particles with masses $m_i$ and
coordinates ${\bf r}_i$. The Jacobi coordinates are defined as
\cite{fed94b,zhu93}.
\begin{eqnarray}\label{ea1}
 {\bf x}_i = {\mu_{jk}} {\bf r}_{jk}, & &
   {\bf y}_i = {\mu_{(jk)i}} {\bf r}_{(jk)i} \nonumber \\
 \mu_{jk} = \left( \frac{1}{m}\frac{m_j m_k}{m_j+m_k} \right)^{1/2}, & &
 \mu_{(jk)i} = \left( \frac{1}{m}\frac{(m_j+m_k)m_i}{m_1+m_2+m_3} \right)^{1/2}\\
 {\bf r}_{jk} = {\bf r}_j-{\bf r}_k, & &
   {\bf r}_{(jk)i} = \frac{m_j {\bf r}_j+m_k {\bf r}_k}{m_j+m_k}-{\bf r}_i \;
 . \nonumber
\end{eqnarray}
where $\{i,j,k\}$ is a cyclic permutation of $\{1,2,3\}$ and $\mu^2$ are
the reduced masses of the subsystems in units of an arbitrary
normalization $m$.

The hyperspherical variables are introduced as
\begin{equation}\label{ea2}
 \rho,\; {\bf n}_{x_i}={\bf x}_i/|{\bf x}_i|,\; 
  {\bf n}_{y_i}={\bf y}_i/|{\bf y}_i|,\; \alpha_i,
\end{equation}
where $\alpha_i$ is in the interval $[0,\pi /2]$
\begin{equation}\label{ea3}
 \rho^2={\bf x_i}^2+{\bf y_i}^2,\; |{\bf x_i}|=\rho\sin\alpha_i,\;
 |{\bf y_i}|=\rho\cos\alpha_i \; .
\end{equation}
We omit the indices where we need not emphasize the particular set of
Jacobi coordinates.  Note that $\rho$ is independent of what set is used.

The relation between three different sets of Jacobi coordinates is given by
\begin{equation} \label{ea5}
{\bf x}_k= {\bf x}_i\cos\varphi_{ik}+ {\bf y}_i\sin\varphi_{ik} ,\;
{\bf y}_k=- {\bf x}_i\sin\varphi_{ik}+ {\bf y}_i\cos\varphi_{ik} \; 
\end{equation}
where the transformation angle $\varphi_{ik}$ is given by the masses as
\begin{equation}  \label{ea7}
\varphi_{ik}=\arctan\left((-1)^p\sqrt{m_j(m_1+m_2+m_3)\over m_k m_i}\right)
\end{equation}
where $(-1)^p$ is the parity of the permutation $\{i,k,j\}$.

\normalsize

\bigskip
\noindent
{\Large \bf Appendix B: B-region solution at intermediate distances}
\setcounter{equation}{0}
\\
\noindent
If the wave function in eq.~(\ref{e67}) is a solution to
eq.~(\ref{e65}) we must have that
\begin{equation}\label{eb1}
 B_+^{II}e^{i \sqrt{\tilde\lambda} \pi/3} +  
 \frac{4}{i \sqrt{3\tilde\lambda}}B_-^{II}
 = \frac{2A_{II}}{\sqrt{3\tilde \lambda}} e^{i \sqrt{\tilde \lambda} \pi/6} \; 
\end{equation}
\begin{equation}\label{eb2}
 \frac{4}{i \sqrt{3\tilde\lambda}}B_+^{II}e^{i \sqrt{\tilde\lambda} \pi/3} - B_-^{II}
 = -  \frac{2A_{II}}{\sqrt{3\tilde\lambda}} e^{i \sqrt{\tilde\lambda} \pi/6} \;
\end{equation}

\begin{equation}\label{eb3}
 ((\kappa_B^{(k)})^2 + v_0\rho^2 + \tilde{\lambda}) B_+^{(k)}e^{\kappa_B^{(k)} \pi/3}
 + \frac{4}{\sqrt{3}}\frac{v_0\rho^2}{\kappa_B^{(k)}}B_-^{(k)} = 0  
\end{equation}
\begin{equation}\label{eb4}
 \frac{4}{\sqrt{3}}\frac{v_0\rho^2}{\kappa_B^{(k)}}  B_+^{(k)}e^{\kappa_B^{(k)} \pi/3}
 - ((\kappa_B^{(k)})^2 + v_0\rho^2 + \tilde{\lambda}) B_-^{(k)} = 0  \;
\end{equation}
for each of the values of k=1,2,3 and furthermore
\begin{eqnarray}\label{eb5}
&  \frac{A_{II}}{2\sqrt{\tilde\lambda}} 
 \left(e^{i \sqrt{\tilde\lambda}(\alpha_0 - \pi/2)} + 
 e^{-i \sqrt{\tilde\lambda}(\alpha_0 - \pi/2)}\right) 
 \nonumber \\ 
& + \frac{1}{i \sqrt{\tilde\lambda}}
 \left(B_+^{II} e^{i \sqrt{\tilde\lambda}(2\pi/3 - \alpha_0)}
 - B_-^{II} e^{-i \sqrt{\tilde\lambda}(2\pi/3 - \alpha_0)} \right) 
 \\
&  + \sum_{k=1}^{3} \left[ \frac{1}{\kappa_B^{(k)}}
 \left(B_+^{(k)} e^{\kappa_B^{(k)} (2\pi/3 - \alpha_0)}
 - B_-^{(k)} e^{-\kappa_B^{(k)} (2\pi/3 - \alpha_0)} \right) \right] 
 + \int_{2\pi/3 -  \alpha_0}^{\alpha_0} \phi_C(\rho,\alpha')d\alpha' = 0 
 \nonumber  .
\end{eqnarray}

Eqs.~(B\ref{eb1}) and (B\ref{eb2}) have for $\lambda \ne 16/3$ the
unique solution
\begin{equation}\label{eb7}
 B_+^{II}e^{i \sqrt{\tilde\lambda} \pi/3} = \frac{1 - 
 \frac{4}{i\sqrt{3\tilde\lambda}}}
 {1 - \frac{16}{3\tilde\lambda}} \; \;
 \frac{2A_{II}}{\sqrt{3\tilde\lambda}} e^{i \sqrt{\tilde\lambda} \pi/6}  \; 
\end{equation}
\begin{equation}\label{eb9}
 B_-^{II} = \frac{1 + \frac{4}{i\sqrt{3\tilde\lambda}}}{1 - 
 \frac{16}{3\tilde\lambda}}\; \;
 \frac{2A_{II}}{\sqrt{3\tilde\lambda}} e^{i \sqrt{\tilde\lambda} \pi/6}  \; .
\end{equation}
For $\tilde\lambda = 16/3$ the two equations determining $B_{\pm}^{II}$ are
identical and infinitely many sets of coefficients exist. The constraint, in
addition to eq.~(B\ref{eb5}), on $B_{\pm}^{II}$ are then
\begin{equation}\label{eb11}
 B_+^{II}e^{i \sqrt{\tilde\lambda} \pi/3} + 
 \frac{4}{i\sqrt{3\tilde\lambda}} B_-^{II} = 
 \frac{2A_{II}}{\sqrt{3\tilde\lambda}} e^{i \sqrt{\tilde\lambda} \pi/6}  \; .
\end{equation}
Eqs.~(B\ref{eb3}) and (B\ref{eb4}) only have non-trivial solutions when
\begin{equation}\label{eb13}
(\kappa_B^{(k)})^2 + v_0\rho^2 + \tilde{\lambda} = 
 \pm  \frac{4i}{\sqrt{3}}\frac{v_0\rho^2}{\kappa_B^{(k)}} \; ,
\end{equation}
which for each of the possible signs is a cubic equation with the
three complex solutions $\kappa_B^{(k)}$ where k=1,2,3. The corresponding
coefficients are then related by
\begin{equation}\label{eb15}
    \pm i B_-^{(k)}  =  B_+^{(k)}e^{\kappa_B^{(k)} \pi/3} \; , \; \; k=1,2,3.
\end{equation}
The solutions corresponding to the different signs are related by an
interchange of the $B_+^{(k)}$ and $B_-^{(k)}$ terms in eq.~(\ref{e67}).

With the expressions in eqs.~(B\ref{eb7}), (B\ref{eb9}) and (B\ref{eb15})
for the coefficients we can rewrite the link to the C-region from
eq.~(B\ref{eb5}) as
\begin{eqnarray}\label{eb17}
 \int_{2\pi/3 -  \alpha_0}^{\alpha_0} \phi_C(\rho,\alpha')d\alpha' = 
 -  \sum_{k=1}^{3}  \left[  \frac{B_-^{(k)}}{\kappa_B^{(k)}}
 \left(\pm i e^{\kappa_B^{(k)} (\pi/3 - \alpha_0)}
  - e^{- \kappa_B^{(k)} (2\pi/3 - \alpha_0)} \right) \right]
  \nonumber \\
 -\frac{A_{II}}{2\sqrt{\tilde\lambda}} \; \frac{1}{1-\frac{16}{3\tilde\lambda}}
  \left( e^{i \sqrt{\tilde\lambda}(\pi/2 - \alpha_0)} 
 (1 + \frac{4}{i \sqrt{3\tilde\lambda}})
  + e^{-i \sqrt{\tilde\lambda}(\pi/2 - \alpha_0)}
  (1 - \frac{4}{i \sqrt{3\tilde\lambda}}) \right) \; .
\end{eqnarray}

Using the wave function $\phi_C$ explicitly together with the
expressions for the C-coefficients from appendix C, we calculate the
integral in eq.~(B\ref{eb5}) and arrive at
\begin{equation}\label{eb19}
 \int_{2\pi/3 -  \alpha_0}^{\alpha_0} \phi_C(\rho,\alpha')d\alpha' =
 \sum_{k=1}^{3} \left[ \frac{\mp i A^{(k)}\sqrt{3}}{\kappa_{AC}^{(k)}}
  \left(e^{\kappa_{AC}^{(k)}(\alpha_0 - \pi/3)}
  -  e^{-\kappa_{AC}^{(k)} (\alpha_0 - \pi/3)} \right) \right] \; ,
\end{equation}
where the summation excluded $k=0$, since this contribution vanishes.
Combined with eq.~(B\ref{eb17}) this gives one constraint between the
coefficients $A_{II}$, $B_-^{(k)}$ and $A^{(k)}$ for k=1,2,3.

\normalsize

\bigskip
\noindent
{\Large \bf Appendix C: A- and C-region solutions at intermediate distances}
\setcounter{equation}{0}\\
\noindent
If the wave functions in eqs.~(\ref{e69}) and (\ref{e71}) are solutions to
eqs.~(\ref{e61}) and (\ref{e63}) we must have that
\begin{equation}\label{ec1}
 ((\kappa_{AC}^{(k)})^2 + v_0\rho^2 + \tilde{\lambda}) A^{(k)}  + 
 \frac{4}{\sqrt{3}}\frac{v_0\rho^2}{\kappa_{AC}^{(k)}} 
 (C_+^{(k)}e^{\kappa_{AC}^{(k)} \pi/3} + C_-^{(k)}e^{-\kappa_{AC}^{(k)} \pi/3}) = 0 \;
\end{equation}
\begin{equation}\label{ec3}
 ((\kappa_{AC}^{(k)})^2 + v_0\rho^2 + \tilde{\lambda}) C_+^{(k)}
e^{\kappa_{AC}^{(k)} \pi/3}
 - \frac{4}{\sqrt{3}}\frac{v_0\rho^2}{\kappa_{AC}^{(k)}} 
 (A^{(k)} +  C_-^{(k)}e^{-\kappa_{AC}^{(k)} \pi/3}) = 0  \;
\end{equation}
\begin{equation}\label{ec5}
 ((\kappa_{AC}^{(k)})^2 + v_0\rho^2 + \tilde{\lambda}) C_-^{(k)}
 e^{-\kappa_{AC}^{(k)} \pi/3}
 - \frac{4}{\sqrt{3}}\frac{v_0\rho^2}{\kappa_{AC}^{(k)}} 
 (A^{(k)} +  C_+^{(k)}e^{\kappa_{AC}^{(k)} \pi/3}) = 0  \;
\end{equation}
for each of the values of k=0,1,2,3 and furthermore
\begin{eqnarray}\label{ec7}
&  \sum_{k=0}^{3} \left[ \frac{A^{(k)}}{\kappa_{AC}^{(k)}}
 \left(e^{\kappa_{AC}^{(k)}(\alpha_0 - \pi/3)}
 +  e^{-\kappa_{AC}^{(k)} (\alpha_0 - \pi/3)} \right) \nonumber
 \right.  \\
& \left. 
  - \frac{1}{\kappa_{AC}^{(k)}} \left(C_+^{(k)} e^{\kappa_{AC}^{(k)}(2\pi/3 - \alpha_0)}
 - C_-^{(k)} e^{-\kappa_{AC}^{(k)} (2\pi/3 - \alpha_0)} \right) \right] 
 + \int_{\alpha_0 - \pi/3}^{2\pi/3 - \alpha_0} \phi_B(\rho,\alpha')d\alpha' 
  = 0 \; .
\end{eqnarray}
Eqs.~(C\ref{ec1}), (C\ref{ec3}) and (C\ref{ec5}) only have non-trivial
solutions when the determinant vanishes for this linear system of
equations. This is equivalent to either 
\begin{equation}\label{ec9}
 (\kappa_{AC}^{(k)})^2 + v_0\rho^2 + \tilde{\lambda} = 0 \;
\end{equation}
or instead, when eq.~(C\ref{ec9}) is false, that
\begin{equation}\label{ec11}
 (\kappa_{AC}^{(k)})^2 + v_0\rho^2 + \tilde{\lambda} = 
 \pm i \frac{4v_0\rho^2}{\kappa_{AC}^{(k)}} \; ,
\end{equation}
The first of these equations in eq.~(C\ref{ec9}) is a second order
equation with two solutions for $\kappa_{AC}^{(k)}$. They only differ by a
sign in $\kappa$ and they therefore correspond to the same
wave function and count as one solution here labeled by $k=0$. The
corresponding coefficients are then related by
\begin{equation}\label{ec13}
  C_+^{(0)} e^{\kappa_{AC}^{(0)} \pi/3} =  - C_-^{(0)} 
 e^{-\kappa_{AC}^{(0)} \pi/3}  = A^{(0)} \;
\end{equation}

The expression in eq.~(C\ref{ec11}) is for each of the possible signs a
cubic equation with three complex solutions $\kappa_{AC}^{(k)}$ here
labeled by k=1,2,3. The corresponding coefficients are given by
\begin{eqnarray}\label{ec15}
 C_+^{(k)} e^{\kappa_{AC}^{(k)} \pi/3} = \frac{A^{(k)}}{2} 
 (1 \mp i \sqrt{3}) \; \; , \; \;
 C_-^{(k)} e^{-\kappa_{AC}^{(k)} \pi/3} = - \frac{A^{(k)}}{2} 
 (1 \pm i \sqrt{3})  \;
\end{eqnarray}
for the values of k=1,2,3. The solutions corresponding to the
different signs are related by an interchange of the $C_+^{(k)}$ and
$C_-^{(k)}$ terms in eq.~(\ref{e71}).

With the expressions in eqs.~(C\ref{ec13}) and (C\ref{ec15}) for the
coefficients we can rewrite the link to the B-region from
eq.~(C\ref{ec7}) as
\begin{eqnarray}\label{ec17}
&  \int_{\alpha_0 - \pi/3}^{2\pi/3 - \alpha_0} \phi_B(\rho,\alpha')d\alpha' 
    \\ 
&  
 = -  \sum_{k=1}^{3} \left[ \frac{3A^{(k)}}{2\kappa_{AC}^{(k)}}
 \left((1 \mp i/ \sqrt{3})e^{\kappa_{AC}^{(k)}(\alpha_0 - \pi/3)}
+ (1 \pm i/ \sqrt{3})e^{-\kappa_{AC}^{(k)} (\alpha_0 - \pi/3)} \right) \right] 
 \nonumber  \; .
\end{eqnarray}
Here the summation does not include $k=0$, since this contribution
vanishes. Using the wave function $\phi_B$ explicitly together with the
expressions for the B-coefficients from appendix B, we calculate the
integral in eq.~(C\ref{ec17}) and arrive at
\begin{eqnarray}\label{ec19}
&  \int_{\alpha_0 - \pi/3}^{2\pi/3 - \alpha_0} \phi_B(\rho,\alpha')d\alpha' 
 = -  \sum_{k=1}^{3} \left(   \frac{B_-^{(k)}(1 \pm i)}{\kappa_B^{(k)}}
 (e^{\kappa_B^{(k)} (\alpha_0 -2\pi/3)} - 
   e^{- \kappa_B^{(k)} (\alpha_0 -\pi/3)})\right) \nonumber \\
& + \left(e^{i \sqrt{\tilde\lambda} (\alpha_0 -2\pi/3)} - 
   e^{- i \sqrt{\tilde\lambda} (\alpha_0 -\pi/3)}\right)
   \frac{4iA_{II}}{\tilde\lambda \sqrt{3}}  \; \; 
 \frac{1}{1 - \frac{16}{3\tilde\lambda}} e^{i \sqrt{\tilde\lambda} \; \pi/6)}
  \; , 
\end{eqnarray}
which combined with eq.~(C\ref{ec17}) gives one constraint between the
coefficients $A_{II}$, $B_-^{(k)}$ and $A^{(k)}$ for k=1,2,3. 

\normalsize

\bigskip
\noindent
{\Large \bf Appendix D: Properties of the asymmetric solutions}
\setcounter{equation}{0}
\\
\noindent
We have not found a general rigorous proof for the claim that three
real solutions exist for $n \ge 3$. However, the numerical
computations all unanimously support the theorem and several very
different limiting cases shall be discussed below. The intermediate
cases are probably similar and with a determined effort possible to
prove. 

When n increases $d_i^2$ approaches zero and 
\begin{equation}\label{ed3}
 S_0  \rightarrow - v_0^{(1)} v_0^{(2)} v_0^{(3)} \; , \; 
 S_1 \rightarrow v_0^{(1)}v_0^{(2)} + v_0^{(1)}v_0^{(3)} + v_0^{(2)}v_0^{(3)}\; , \;
 S_2 \rightarrow - v_0^{(1)} -v_0^{(2)} - v_0^{(3)} \; \;
\end{equation}
with the solution $\epsilon = v_0^{(i)}$ to eq.~(\ref{e125}).

In general the cubic equation in eq.~(\ref{e125}) has three real solutions if
\begin{equation}\label{ed1}
 \left( \frac{1}{3}S_1-\frac{1}{9}S_2^2 \right)^3
  + \left( -\frac{1}{6} S_1 S_2 + \frac{1}{2} S_0 + \frac{1}{27}S_2^3 \right) ^2
  \le 0  \;  ,
\end{equation}
which can be rewritten as
\begin{equation}\label{ed5}
 \frac{1}{27}S_1^3 - \frac{1}{108} S_1^2 S_2^2 + \frac{1}{4} S_0^2
  + \frac{1}{27} S_0 S_2^3  - \frac{1}{6} S_0 S_1 S_2 \le 0 \; .
\end{equation}
When one of the potentials, for example $v_0^{(3)}$, is very small or
vanishes we get in this limit
\begin{equation}\label{ed7}
 S_0 \rightarrow 0 \; ,\;
 S_1 \rightarrow  v_0^{(1)}v_0^{(2)} (1- d_3^2) \; ,\;
 S_2 \rightarrow  - v_0^{(1)} -v_0^{(2)} \;
\end{equation}
and the condition in eq.~(D\ref{ed5}) becomes
\begin{equation}\label{ed9}
 \frac{1}{27} S_1^3 - \frac{1}{108} S_1^2 S_2^2 = 
 - \frac{1}{27} \left(\frac{1}{4} (v_0^{(1)} -v_0^{(2)})^2 +  
 v_0^{(1)}v_0^{(2)} d_3^2\right) \le - \frac{1}{27} (v_0^{(1)})^2 (1-d_3^2) \le 0 \; .
\end{equation}
When all $v_0^{(i)}$ are equal the condition in eq.~(D\ref{ed5}) reduces to 
\begin{equation}\label{ed11}
 \frac{1}{27} x^3 - \frac{1}{12} x^2  + \frac{1}{4} y^2
  -  y  - \frac{1}{2} x y \le 0 \; ,
\end{equation}
where the overall factor $(v_0^{(i)})^6$ has been removed and the new
variables $x$ and $y$ are defined as
\begin{equation}\label{ed13}
 x =  3 - d_1^2 - d_2^2 - d_3^2 \; , \; \;
 y = 2d_1 d_2 d_3 + d_1^2 + d_2 ^2 + d_3^2 - 1  \; .
\end{equation}
Direct computation of the quantities reformulates eq.~(D\ref{ed11})
into
\begin{equation}\label{ed15}
 - \frac{1}{27} (d_1^2 + d_2^2 + d_3^2)^3 + (d_1 d_2 d_3)^2  \le 0 \; ,
\end{equation}
which is fulfilled for any set of $d_i$.

\normalsize

\bigskip
\noindent
{\Large \bf Appendix E: Eigenvalue equation at large distances for three different particles}
\setcounter{equation}{0}
\\
\noindent
The matching conditions at $\alpha_0^{(1)}, \alpha_0^{(2)}$ and
$\alpha_0^{(3)}$ at large distances for the asymmetric case provide
the eigenvalue equation. This is found by equating the functions in
eqs.~(\ref{e133}) and (\ref{e135}) and their first derivatives, i.e.\
\begin{eqnarray}\label{ee1}
 & \left[a_i \sin \left((\alpha_0^{(i)} - \pi/2) \sqrt{\tilde\lambda}\right)
  - c_i \sin \left(\alpha_0^{(i)}  \sqrt{\tilde\lambda}\right) \right]
  = b_i \sin (\alpha_0^{(i)} \kappa_i)   \\
 & \left[a_i \cos \left((\alpha_0^{(i)} - \pi/2) \sqrt{\tilde\lambda}\right) 
 - c_i \cos \left(\alpha_0^{(i)}  \sqrt{\tilde\lambda}\right)\right]
  \sqrt{\tilde\lambda}
  = b_i \kappa_i \cos (\alpha_0^{(i)} \kappa_i)   
\end{eqnarray}
and then eliminating $b_i$ resulting in
\begin{eqnarray}\label{ee3}
 & \left[a_i \sin \left((\alpha_0^{(i)} - \pi/2) \sqrt{\tilde\lambda}\right)
  - c_i \sin \left(\alpha_0^{(i)}  \sqrt{\tilde\lambda}\right) \right]
  \kappa_i  \cos(\alpha_0^{(i)}  \kappa_i) = \nonumber \\
 & \left[a_i \cos \left((\alpha_0^{(i)} - \pi/2) \sqrt{\tilde\lambda}\right)
  - c_i \cos \left(\alpha_0^{(i)}  \sqrt{\tilde\lambda}\right) \right]
  \sqrt{\tilde\lambda}  \sin(\alpha_0^{(i)} \kappa_i) \;  .
\end{eqnarray}
Furthermore eliminating $c_i$ by use of eqs.~(\ref{e137})-(\ref{e136})
then provide 3 linear equations in $a_i$. They only have non-trivial
solutions when the corresponding determinant, $D = $det$ \{d_{ik}\}$
vanishes. The matrix elements are
\begin{eqnarray}\label{ee5}
  d_{ii} = \kappa_i \sin \left((\alpha_0^{(i)} - \pi/2) 
 \sqrt{\tilde\lambda}\right)    \cos(\alpha_0^{(i)} \kappa_i)
  - \sqrt{\tilde\lambda}
    \cos \left((\alpha_0^{(i)} - \pi/2) \sqrt{\tilde\lambda}\right)
    \sin (\alpha_0^{(i)} \kappa_i)
\end{eqnarray}    
\begin{equation}\label{ee7}
   d_{ik} = \frac{A_i f_j}{F}  \; , \; {\rm for} \; \; i\neq  k \;  , 
\end{equation}
where $f_j$ is defined in eq.~(\ref{e136}) and
\begin{eqnarray}\label{ee9}
   A_i = & \frac{2F}{\sqrt{\tilde\lambda}} 
  \left[ \kappa_i
  \sin \left(\alpha_0^{(i)}  \sqrt{\tilde\lambda}\right) 
 \cos(\alpha_0^{(i)} \kappa_i) 
 - \sqrt{\tilde\lambda}
   \cos \left(\alpha_0^{(i)}  \sqrt{\tilde\lambda}\right) 
 \sin(\alpha_0^{(i)} \kappa_i)
 \right]
\end{eqnarray}    
By further defining
\begin{equation}\label{ee11} 
  B_i = \frac{d_{ii} f_i^2}{F^2}
\end{equation}
the determinant can then be written as
\begin{equation}\label{ee13}
 D = B_1 B_2 B_3 + 2A_1 A_2 A_3 - B_1 A_2 A_3  - A_1 A_2 B_3 -  A_1 B_2 A_3 
  \; .
\end{equation}
The eigenvalues $\tilde\lambda$ are then determined by $D = 0$.

\normalsize

\bigskip
\noindent
{\Large \bf Appendix F: Solutions to the case of two identical
spin-1/2 particles} 
\setcounter{equation}{0} \\
\noindent
The solutions to the Faddeev equations in eqs.~(\ref{e174}) and
(\ref{e176}) are first found independently for each of the four
regions of fig.4. We shall use the notation $\alpha_{ff}=\alpha_{1}$
and $\alpha_{fc}= \alpha_{2}$.  In region II , where all potentials
are zero we have
\begin{equation}\label{ef1}
  \phi^{(i)}_s(\alpha_{i}) = 
        A_i \sin\left((\alpha_{i}-\pi/2)\sqrt{\tilde{\lambda}}\right) \; , 
\end{equation}
where $(i,s)=(1,0),(2,{s_c-1/2}),(3,{s_c+1/2})$ and eq.~(\ref{e173})
relate these to the remaining spin components of $\phi^{(3)}$ and
$\phi^{(2)}$.

In region B of fig.4, where only one of the potentials is identically
zero, we have instead
\begin{eqnarray}\label{ef3}
 \phi^{(1)}_0(\alpha_{ff}) &=& 
            b_1 \sin(\alpha_{ff} \kappa_{0})
  + a_0 \sin(\alpha_{ff} \sqrt{\tilde{\lambda}}) \\ 
 \phi^{(i)}_s(\alpha_{fc}) &=& 
        b_i \sin\left((\alpha_{fc}-\pi/2)\sqrt{\tilde{\lambda}}\right)  \; ,
\end{eqnarray}
where $(i,s)=(2,{s_c-1/2}),(3,{s_c+1/2})$, $\kappa_{0}=\sqrt{v^{(ff)}_0
\rho^2 +\tilde{\lambda}(\rho)}$ and
\begin{equation}\label{ef5}
  a_0 =   - \frac{4f}{\sqrt{\tilde{\lambda}}}  
   (C_{0,s_c-1/2}^{1 2} A_2 - C_{0,s_c+1/2}^{1 2} A_3) \; , 
\end{equation}
where we in analogy to eq.~(\ref{e136}) define
\begin{equation}\label{ef4}
 f = \frac{\sin\left((\varphi-\pi/2)\sqrt{\tilde{\lambda}}\right)}
   {\sin(2\varphi)} \; , \;
 \tilde{f} = \frac{\sin\left((\tilde\varphi-\pi/2)\sqrt{\tilde{\lambda}}
 \right)}  {\sin(2\tilde\varphi)} \; .
\end{equation}

In region C of fig.4, where the other potential is identically zero,
we have
\begin{eqnarray} \label{ef7}
 \phi^{(1)}_0(\alpha_{ff}) &=&
          c_1 \sin\left((\alpha_{ff}-\pi/2)\sqrt{\tilde{\lambda}}\right)  \\
 \phi^{(i)}_s(\alpha_{fc}) &=&
          c_i \sin(\alpha_{fc} \kappa_{s})               
 + a_s \sin(\alpha_{fc} \sqrt{\tilde{\lambda}}) \; ,
\end{eqnarray}
where $(i,s)=(2,{s_c-1/2}),(3,{s_c+1/2})$, $\kappa_{s_c \pm
1/2}=\sqrt{v^{(fc\pm)}_0 \rho^2 +\tilde{\lambda}(\rho)}$ and
\begin{eqnarray} \label{ef9}
 a_{s_c-1/2} =  -\frac{2}{\sqrt{\tilde{\lambda}}}  
 \left[ C_{0,s_c-1/2}^{12} A_1 f
+ (C_{s_c-1/2,s_c-1/2}^{23} A_2 + C_{s_c-1/2,s_c+1/2}^{23} A_3) \tilde{f}
                                                  \right]  \\
 a_{s_c+1/2} = -\frac{2}{\sqrt{\tilde{\lambda}}} 
 \left[
  c_{0,s_c+1/2}^{12} A_1 f
+ (-C_{s_c-1/2,s_c+1/2}^{23} A_2 + C_{s_c+1/2,s_c+1/2}^{23} A_3) \tilde{f}
                                                  \right] \; .
\end{eqnarray}

In region D of fig.4, where both potentials are finite, we have
\begin{equation} \label{ef11}
  \phi^{(i)}_s(\alpha_{i}) =
            d_i \sin(\alpha_{i} \kappa_{s})
      + a_s \sin(\alpha_{i} \sqrt{\tilde{\lambda}})  \; ,
\end{equation}
where $(i,s)=(1,0),(2,{s_c-1/2}),(3,{s_c+1/2})$.

Imposing continuity of the functions and their first derivatives at
the boundaries between the different regions of fig.4, the constants
must be related by
\begin{eqnarray} \label{ef13}
 c_1  =  A_1 \;, \; b_2  =  A_2 \;, \;  b_3  =  A_3 \;, \;  
 d_1  =  b_1 \;, \;  d_2  =  c_2 \;, \;  d_3  =  c_3   \;, \; 
\end{eqnarray}
and the remaining six constants obey the following set of six linear
equations:
\begin{eqnarray} \label{ef15}
& A_1 \sin\left( (\alpha^{(ff)}_0-\frac{\pi}{2}) \sqrt{\tilde{\lambda}}
\right) = b_1 \sin(\alpha^{(ff)}_0 \kappa_{0})  
+ a_0 \sin(\alpha^{(ff)}_0 \sqrt{\tilde{\lambda}}) \nonumber \\
& A_1 \sqrt{\tilde{\lambda}} \cos\left( (\alpha^{(ff)}_0-\frac{\pi}{2}) 
\sqrt{\tilde{\lambda}} \right)
= b_1 \kappa_{0} \cos(\alpha^{(ff)}_0 \kappa_{0}) 
 + a_0 \sqrt{\tilde{\lambda}} \cos(\alpha^{(ff)}_0 \sqrt{\tilde{\lambda}}) 
 \nonumber \\
& A_2 \sin\left( (\alpha^{(fc)}_0-\frac{\pi}{2}) \sqrt{\tilde{\lambda}} \right)
= c_2 \sin(\alpha^{(fc)}_0 \kappa_{s_c-1/2}) 
  +a_{s_c-1/2}   \sin(\alpha^{(fc)}_0 \sqrt{\tilde{\lambda}}) \\
& A_2 \sqrt{\tilde{\lambda}}\cos\left( (\alpha^{(fc)}_0-\frac{\pi}{2})
\sqrt{\tilde{\lambda}} \right) = c_2 \kappa_{s_c-1/2}
\cos(\alpha^{(fc)}_0 \kappa_{s_c-1/2}) - a_{s_c-1/2} 
 \sqrt{\tilde{\lambda}} \cos(\alpha^{(fc)}_0 \sqrt{\tilde{\lambda}}) \nonumber \\
& A_3 \sin\left( (\alpha^{(fc)}_0-\frac{\pi}{2}) \sqrt{\tilde{\lambda}}
\right)
= c_3 \sin(\alpha^{(fc)}_0 \kappa_{s_c+1/2}) 
+ a_{s_c+1/2} \sin(\alpha^{(fc)}_0 \sqrt{\tilde{\lambda}}) \nonumber \\
& A_3 \sqrt{\tilde{\lambda}}\cos\left( (\alpha^{(fc)}_0-\frac{\pi}{2})
\sqrt{\tilde{\lambda}} \right) = c_3 \kappa_{s_c+1/2}
\cos(\alpha^{(fc)}_0 \kappa_{s_c+1/2}) + a_{s_c+1/2} 
 \sqrt{\tilde{\lambda}} \cos(\alpha^{(fc)}_0 \sqrt{\tilde{\lambda}}) 
 \nonumber  \; .
\end{eqnarray}

Since $a_s$ only depend on $A_i$ we can easily eliminate $b_1, c_2,
c_3$ from the equations in eq.~(F\ref{ef15}). This leaves 3 homogeneous
linear equations in $A_1, A_2, A_3$ with a corresponding determinant
$D=$det$\{d_{ik}\}$. The matrix elements are given by
\begin{eqnarray} \label{ef19}
 d_{22} = D_2 + F_{23} C_{s_c-1/2,s_c-1/2}^{23}  \; , \; 
 d_{33} = D_3 +  F_{32} C_{s_c+1/2,s_c+1/2}^{23}  \; ,  \\
 d_{11} = D_1 \; , \; d_{ik} = F_{ik} C_{s_i,s_k}^{ik} \; ,
\end{eqnarray} 
where $s_1=0, s_2= s_c-1/2, s_3 =s_c +1/2$, the spin overlap functions
are defined in eqs.~(\ref{e29}) and
\begin{equation} \label{ef21}
 D_i = \kappa_{s}
   \sin\left( (\alpha^{(i)}_0-\frac{\pi}{2}) \sqrt{\tilde{\lambda}} \right)
       \cos(\alpha^{(i)}_0 \kappa_{s})
      - \sqrt{\tilde{\lambda}}
       \cos\left( (\alpha^{(i)}_0-\frac{\pi}{2}) \sqrt{\tilde{\lambda}} \right)
       \sin(\alpha^{(i)}_0 \kappa_{s}) \; ,
\end{equation} 
\begin{eqnarray} \label{ef23}
  F_{ik} = \left[\kappa_{s}
   \sin\left( \alpha^{(i)}_0 \sqrt{\tilde{\lambda}} \right)
       \cos(\alpha^{(i)}_0 \kappa_{s})
      - \sqrt{\tilde{\lambda}}
       \cos\left( \alpha^{(i)}_0 \sqrt{\tilde{\lambda}} \right)
       \sin(\alpha^{(i)}_0 \kappa_{s}) \right]   
   \frac{2(1 + \delta_{i,1})}{\sqrt{\tilde{\lambda}}} f_j \; ,
\end{eqnarray} 
where $f_j$ is defined in eq.~(\ref{e136}).

The values of $\tilde{\lambda}$ are as usual determined from $D=0$.

Simplified expressions can be obtained in the limit of very large
distances where
\begin{equation}\label{ef25}
 F_{ik} \approx -2(1 + \delta_{i,1}) \mu_{jk} \sqrt{v_0^{(i)}} a_{scat}^{(i)} 
  \cos(X_i\sqrt{v_0^{(i)}}) f_j \; ,
\end{equation} 
\begin{equation}\label{ef27}
 D_i \approx - \cos(X_i\sqrt{v_0^{(i)}}) \sqrt{v_0^{(i)}}
 \left(\rho \sin(\frac{\pi}{2}\sqrt{\tilde\lambda}) + 
  a_{scat}^{(i)} \mu_{jk} \sqrt{\tilde\lambda}
  \cos(\frac{\pi}{2}\sqrt{\tilde\lambda}) \right)  \; .
\end{equation}
Here $i=1,2,3$ corresponds to $ff, fc-, fc+$, respectively. Then
$v_0^{(i)}$ is defined in eqs.~(\ref{e185}) and (\ref{e194}) and the
scattering lengths, $a_{scat}^{(ff)}$, $a_{scat}^{(fc+)}$,
$a_{scat}^{(fc-)}$, for the two-body systems are defined in analogy with
eq.~(\ref{e157}) in terms of radii, potentials and reduced masses.

When $\tilde{\lambda} /\rho^2$ remains finite the dominating terms in
the determinant are $D_i$. Thus $D$ is approximately diagonal with the
matrixelemnts $D_i$ given in eq.~(F\ref{ef21}). The eigenvalues in the
limit of large $\rho$ then again corresponds to the two-body bound
states obtained by using eq.~(F\ref{ef21}) and solving $D_i=0$.


\begin{thebibliography}{99}
\bibitem{fed93}D.~V.~Fedorov and A.~S.~Jensen, Phys.~Rev.~Lett. {\bf
71}, 4103 (1993).

\bibitem{fed94b} D.V.~Fedorov, A.S.~Jensen and K.~Riisager, Phys.~Rev.
  {\bf C50}, 2372 (1994).

 \bibitem{mac68} J.H.~Macek, J.Phys. {\bf B1}, 831 (1968).

\bibitem{efi70} V.N.~Efimov, Phys.~Lett. {\bf B33}, 563 (1970) ;
  Sov.Journ.Nucl.Phys. {\bf 12}, 589 (1971).

\bibitem{cor86} Th.~Cornelius and W.~Gl\"{o}ckle, J.~Chem.~Phys.
  {\bf 85}, 3906 (1986).

\bibitem{fed94c}D.V.~Fedorov, A.S.~Jensen and K.~Riisager, Phys.~Rev.~Lett.
 {\bf 73}, 2817 (1994).

\bibitem{ric92} J.-M. Richard, Phys. Reports {\bf 212}, 1 (1992).

 \bibitem{zhu93}M.V.~Zhukov, B.V.~Danilin, D.V.~Fedorov, J.M.~Bang,
   I.J.~Thompson and J.S.~Vaagen,  Phys. Reports {\bf 231}, 151 (1993).

\bibitem{fed94a} D.V.~Fedorov, A.S.~Jensen and K.~Riisager, Phys.~Rev.
  {\bf C49}, 201 (1994).

\bibitem{coo93} A.R.~Cooper, S.~Jain and J.M.~Hutson, J.~Chem.~Phys.
  {\bf 98}, 2160 (1993).

\bibitem{jac94} J.D.~Jackson, Phys.~Rev {\bf A49}, 132 (1994).

\bibitem{ros91} J.M~Rost and J.~S.~Briggs, J. of Phys. {\bf B24}, 4293 (1991).

\bibitem{lin95} C.D.~Lin, Phys. Reports {\bf 257}, 1 (1995).

 \bibitem{efi90}V.~Efimov, Comm. Nucl. Part. Phys. {\bf 19}, 271 (1990).

\bibitem{kie94} A.~Kievsky, M.~Viviani and S.~Rosati, Nucl. Phys~{\bf
A577}, 511 (1994).

\bibitem{fri95} J.L.~Friar,  XIV'th Int. Conf. on Few-Body Problems 
in Physics, Williamsburg, AIP Conference Proceedings {\bf 334}, 323 (1995).

\bibitem{han95}P.G.~Hansen, A.S.~Jensen and B.~Jonson, 
 Ann.~Rev.~Nucl.~Part.~Sci. {\bf 45}, 591 (1995). 

 \bibitem{bar92} N.~Barnea and V.B.~Mandelzweig, Phys.~Rev.
  {\bf C45}, 1458 (1992).

\bibitem{fed95} D.V.~Fedorov, E.~Garrido and A.S.~Jensen, Phys.~Rev.
  {\bf C51}, 3052 (1995).


\end{thebibliography}
\end{document}